\documentclass{aa}
\usepackage[varg]{txfonts}
\usepackage{graphicx}
\usepackage{natbib}
\usepackage{subfig}
\usepackage{hyperref}

\bibpunct{(}{)}{;}{a}{}{,} 

\begin{document}
%
%
    \title{A dynamical transition from atomic to molecular\\ intermediate-velocity clouds \thanks{Based on observations obtained with Planck (http://www.esa.int/Planck), an ESA science mission with instruments and contributions directly funded by ESA Member States, NASA, and Canada.}}
    \titlerunning{Atomic and molecular intermediate-velocity clouds}


   \author{T.~R{\"o}hser \inst{1} 
	   \and
           J.~Kerp \inst{1}
	   \and 
	   B.~Winkel \inst{2}
	   \and
	   F.~Boulanger \inst{3}
	   \and
	   G.~Lagache \inst{3}
           }

   \institute{Argelander-Institut f{\"u}r Astronomie (AIfA), Universit{\"a}t Bonn, Auf dem H{\"u}gel 71, D-53121 Bonn\\
              \email{troehser@astro.uni-bonn.de}
              \and
              Max-Planck-Institut f{\"u}r Radioastronomie (MPIfR), Auf dem H{\"u}gel 69, D-53121 Bonn
	      \and
	      Institut d'Astrophysique Spatiale (IAS), Universit\'{e} Paris-Sud XI, F-91405 Orsay
              }

   \date{Received ---; accepted ---}

    \abstract
    {Towards the high galactic latitude sky, the far-infrared (FIR) intensity is tightly correlated to the total hydrogen column density which is made up of atomic ($\ion{H}{i}$) and molecular hydrogen (H$_{2})$. Above a certain column density threshold, atomic hydrogen turns molecular.}
    {We analyse gas and dust properties of intermediate-velocity clouds (IVCs) in the lower galactic halo to explore their transition from the atomic to the molecular phase. Driven by observations, we investigate the physical processes that transform a purely atomic IVC into a molecular one.}
    {Data from the Effelsberg-Bonn $\ion{H}{i}$-Survey (EBHIS) are correlated to FIR wavebands of the Planck satellite and IRIS. Modified black-body emission spectra are fitted to deduce dust optical depths and grain temperatures. We remove the contribution of atomic hydrogen to the FIR intensity to estimate molecular hydrogen column densities.}
    {Two IVCs show different FIR properties, despite their similarity in $\ion{H}{i}$, such as narrow spectral lines and large column densities. One FIR bright IVC is associated with H$_{2}$, confirmed by $^{12}$CO $(1\rightarrow0)$ emission; the other IVC is FIR dim and shows no FIR excess, which indicates the absence of molecular hydrogen.}
    {We propose that the FIR dim and bright IVCs probe the transition between the atomic and molecular gas phase. Triggered by dynamical processes, this transition happens during the descent of IVCs onto the galactic disk. The most natural driver is ram pressure exerted onto the cloud by the increasing halo density. Because of the enhanced pressure, the formation timescale of H$_{2}$ is reduced, allowing the formation of large amounts of H$_{2}$ within a few Myr.}


   \maketitle

\section{Introduction}
\label{sec:introduction}

The Infrared Astronomical Satellite (IRAS) showed that, up to a certain threshold, the $\ion{H}{i}$ $21\,\mathrm{cm}$ emission correlates linearly with the far-infrared (FIR) dust continuum at high galactic latitudes \citep{Low1984,Boulanger1988,Boulanger1996}. Despite this linear relationship, previous studies towards high galactic latitudes \citep[e.g.][]{Desert1988,Reach1998} find excess FIR radiation associated with molecular gas, which is traced by carbon monoxide (CO) emission \citep{Magnani1985}. Several of these objects show radial velocities relative to the local standard of rest (LSR) between $30\,\mathrm{km}\,\mathrm{s}^{-1} \leq |v_{\mathrm{LSR}}| \leq 90\,\mathrm{km}\,\mathrm{s}^{-1}$ that are difficult to account for with a simple model of galactic rotation. These clouds are classified as intermediate-velocity clouds (IVCs) and, as a subcategory, intermediate-velocity molecular clouds (IVMCs).

Intermediate-velocity clouds are thought to originate from a galactic fountain process \citep{Bregman2004}. Typically they show up with metal abundances close to solar, a traceable dust content, and distances below $5\,\mathrm{kpc}$ \citep{Wakker2001}. Commonly, IVCs emit in the FIR \citep[e.g.][]{Planckcollaboration2011XXIV}. Towards many IVCs a diffuse H$_{2}$ column density of $N_{\mathrm{H}_{2}} = 10^{14}-10^{16}\,\mathrm{cm}^{-2}$ is inferred by interstellar absorption line measurements \citep{Richter2003,Wakker2006}; however, only a few IVMCs are known \citep{Magnani2010}. Dynamical processes resulting from the motion of IVCs through the halo and their descent onto the galactic disk are thought to play a major role in the process of H$_{2}$ formation in IVCs \citep{Odenwald1987,Desert1990a,Gillmon2006b,Guillard2009}. The transition between the atomic and the molecular phase happens at a total hydrogen column density of $N_{\mathrm{H}} = 2.0-5.0 \times 10^{20}\,\mathrm{cm}^{-2}$ \citep{
Savage1977,Reach1994,Lagache1998,Gillmon2006a,Gillmon2006b,Planckcollaboration2011XXIV}. 

In this paper we study the gas and dust properties of IVCs and their transition from the atomic to the molecular phase by correlating the $\ion{H}{i}$ distribution to the FIR dust emission at high galactic latitudes. We are interested in the conditions required for the transition from $\ion{H}{i}$ to H$_{2}$ in IVCs at the disk-halo interface.

We use data from the new Effelsberg-Bonn $\ion{H}{i}$-Survey \citep[EBHIS,][]{Winkel2010,Kerp2011}, the Planck Satellite \citep{Planckcollaboration2013I}, and IRIS \citep{MivilleDeschenes2005b}. Our field of interest has a size of $25 \times 25\,\mathrm{deg}^{2}$ centred on galactic coordinates ($l$, $b$) $=$ ($235^{\circ}$, $65^{\circ}$). This area is located between the Intermediate-Velocity (IV) Arch and IV Spur which are two large structures in the distribution of IVCs \citep{Wakker2004}. In \citet{Planckcollaboration2013XXX} this field is used also in the analysis of the cosmic infrared background.

Section \ref{sec:data} gives the properties of our data in $\ion{H}{i}$ and the FIR. Section \ref{sec:methods} describes the $\ion{H}{i}$-FIR correlation in more detail. Section \ref{sec:entire-field} presents the observational results inferred from the $\ion{H}{i}$, the FIR, and their correlation for the entire field, while Sect. \ref{sec:individual-clouds} gives results for two particular IVCs within the field. Section \ref{sec:discussion} discusses a dynamically driven $\ion{H}{i}$-H$_{2}$ transition with respect to the two IVCs and Sect. \ref{sec:conclusion} summarises our results.

\section{Data}
\label{sec:data}

Table \ref{tab:data} gives the main characteristics of the different data sets. In the FIR we use data at $353$, $545$, and $857\,\mathrm{GHz}$ from Planck and $3000\,\mathrm{GHz}$ from IRIS. These frequencies probe the peak of the FIR dust emission.

The Effelsberg-Bonn $\ion{H}{i}$-Survey \citep[EBHIS,][]{Winkel2010,Kerp2011} is a new, fully sampled survey of the northern hemisphere in $\ion{H}{i}$ $21\,\mathrm{cm}$ line emission above declinations of $-5\,^{\circ}$. FPGA-based spectrometers \citep{Stanko2005} allow one to conduct in parallel a galactic and extragalactic survey out to redshifts of $z \simeq 0.07$. The angular resolution is approximately $10.8\,\arcmin$, and with a channel width of $1.3\,\mathrm{km}\,\mathrm{\mathrm{s}}^{-1}$ the rms noise is less than $90\,\mathrm{mK}$. The data is corrected for stray radiation which is especially important for studies of the diffuse ISM at high galactic latitudes. 

The Planck satellite covers the spectral range between $25$ and $1000\,\mathrm{GHz}$ \citep{Planckcollaboration2013I}. The High Frequency Instrument (HFI) has bolometer detectors that offer six bands centred on $100$, $143$, $217$, $353$, $545$, and $857\,\mathrm{GHz}$.

We use the Planck maps that are corrected for zodiacal emission \citep{Planckcollaboration2013XIV}, but no CMB emission is subtracted. The $353\,\mathrm{GHz}$ data is transformed from the original $K_{\mathrm{CMB}}$ units to $\mathrm{MJy}\,\mathrm{sr}^{-1}$ by applying the conversion factor given in \citet{Planckcollaboration2013IX}. No colour correction is done for any FIR frequency.

Infrared dust continuum data is available from the Infrared Astronomical Satellite (IRAS) at $12$, $25$, $60$, and $100\,\mathrm{\mu m}$ \citep{Neugebauer1984}. The four IRAS wavelengths have been revised in the Improved Reprocessing of the IRAS Survey \citep[IRIS,][]{MivilleDeschenes2005b}. They preserve the full angular resolution and perform a thorough absolute calibration. In recent data products the stripe of missing data in IRAS is filled with observations from DIRBE of lower angular resolution.

\begin{table}[!t]
  \caption{Properties of the $\ion{H}{i}$ and FIR data. The columns list (from left to right) the survey, the corresponding frequency, the angular resolution, and the average noise.}  
  \label{tab:data}
  \small
  \centering
  \begin{tabular}{ccccc}
    \hline\hline
    Survey & $\nu$ [$\mathrm{GHz}$] & FWHM [$\arcmin$] & Average noise & References \\ 
    \hline
    EBHIS & $1.42$ & $10.8$ & $\lesssim 90\,\mathrm{mK}$ & 1 \\
    Planck & $353$ & $4.82$ & $0.012\,\mathrm{MJy}\,\mathrm{sr}^{-1}$ & 2 \\
    & $545$ & $4.68$ & $0.015\,\mathrm{MJy}\,\mathrm{sr}^{-1}$ & 2 \\
    & $857$ & $4.33$ & $0.016\,\mathrm{MJy}\,\mathrm{sr}^{-1}$ & 2 \\
    IRIS & $3000$ & $4.3$ & $0.06\,\mathrm{MJy}\,\mathrm{sr}^{-1}$ & 3 \\
    \hline
  \end{tabular}
  \tablebib{
  (1) \citet{Winkel2010}; (2) \citet{Planckcollaboration2013I}; (3) \citet{MivilleDeschenes2005b}
  }
\end{table}

\section{Methods}
\label{sec:methods}

We correlate the $\ion{H}{i}$ column density $N_{\ion{H}{i}}$ to the FIR intensity $I_{\nu}$ at frequency $\nu$. The quantitative correlation of two different data sets like $\ion{H}{i}$ and the FIR demands a common coordinate grid with the same angular resolution. Hence we convolve the FIR maps with two-dimensional Gaussian functions that have a spatial width given by the full width half maximum (FWHM) fulfilling $\mathrm{FWHM}_{\mathrm{\ion{H}{i}}}^{2} =\mathrm{FWHM}_{\mathrm{Gauss}}^{2}+\mathrm{FWHM}_{\mathrm{FIR}}^{2}$.

At high galactic latitudes the $\ion{H}{i}$-FIR correlation is linear \citep[e.g.][]{Boulanger1996} and it can be written as
\begin{equation}
\label{eq:1}
  I_{\nu} = R_{\nu} + \epsilon_{\nu} \times N_{\ion{H}{i}},
\end{equation}
where $\epsilon_{\nu}$ is the dust emissivity per nucleon. A general offset $R_{\nu}$ accounts for emission that is not  associated with the galactic gas distribution, for example  the cosmic infrared background. The values of each pixel of the $\ion{H}{i}$ and FIR maps are plotted against each other. With standard least-squares techniques the linear parameters and their statistical errors are estimated.

In order to exclude point sources from the fit, we apply a first fit to reject all data points that deviate by more than $3\sigma$ from this initial estimate. In a second step, we reject all pixels in which more $\ion{H}{i}$ emission is in IVCs between $-60\,\mathrm{km}\,\mathrm{s}^{-1} \leq v_{\mathrm{LSR}} \leq -20\,\mathrm{km}\,\mathrm{s}^{-1}$ than in the local gas with $-20\,\mathrm{km}\,\mathrm{s}^{-1} \leq v_{\mathrm{LSR}} \leq +20\,\mathrm{km}\,\mathrm{s}^{-1}$. We do this because of the generally lower dust emissivities of IVCs compared to local gas \citep{Planckcollaboration2011XXIV}, which is expected to bias the fit.

We only quote errors on the emissivities that represent statistical uncertainties. These uncertainties are lower than the true errors \citep{Planckcollaboration2011XXIV} since we do not consider effects like the cosmic infrared background or residual zodiacal emission.

Equation \eqref{eq:1} can be generalised by adding more $\ion{H}{i}$ components with different dust emissivities. As \citet{Peek2009} show, this superposition can easily produce false estimates for a FIR dim cloud behind a bright local foreground. In this paper we concentrate on two bright IVCs which completely dominate the emission along their lines of sight. Thus, a superposition of several components is not required in order to model their FIR brightness.

The emissivity $\epsilon_{\nu}$ in Eq. \eqref{eq:1} depends on the dust-to-gas ratio, but also on gas and dust properties. Deviations from a linear behaviour are mostly due to local variations of the amount of neutral atomic gas where significant amounts of the hydrogen are in either ionised or molecular forms. These species will be missing from the $\ion{H}{i}$-FIR correlation. Above an empirical threshold of $N_{\ion{H}{i}} = 2 \times 10^{20}\,\mathrm{cm}^{-2}$, molecular hydrogen steepens the correlation \citep[e.g.][]{Planckcollaboration2011XXIV}, which can be inferred without tracer molecules like CO \citep{Reach1998}. This steepening is not due to $\ion{H}{i}$ absorption, since opacity corrections are relevant for $N_{\ion{H}{i}} > 10^{21}\,\mathrm{cm}^{-2}$ \citep{Strasser2004} which is much more than the column density observed with EBHIS in galactic cirrus clouds. 

\begin{figure*}[!t]
 \centering
  \subfloat{\includegraphics[width=0.5\textwidth]{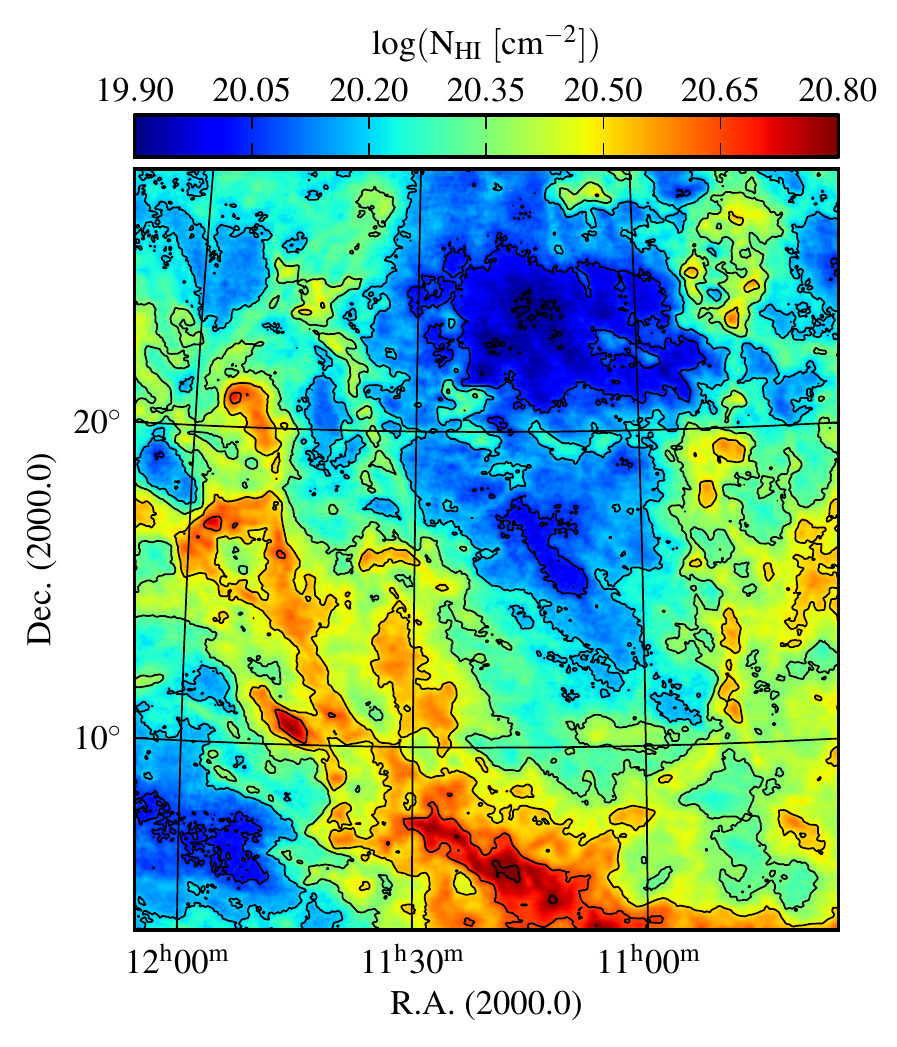}} \subfloat{\includegraphics[width=0.5\textwidth]{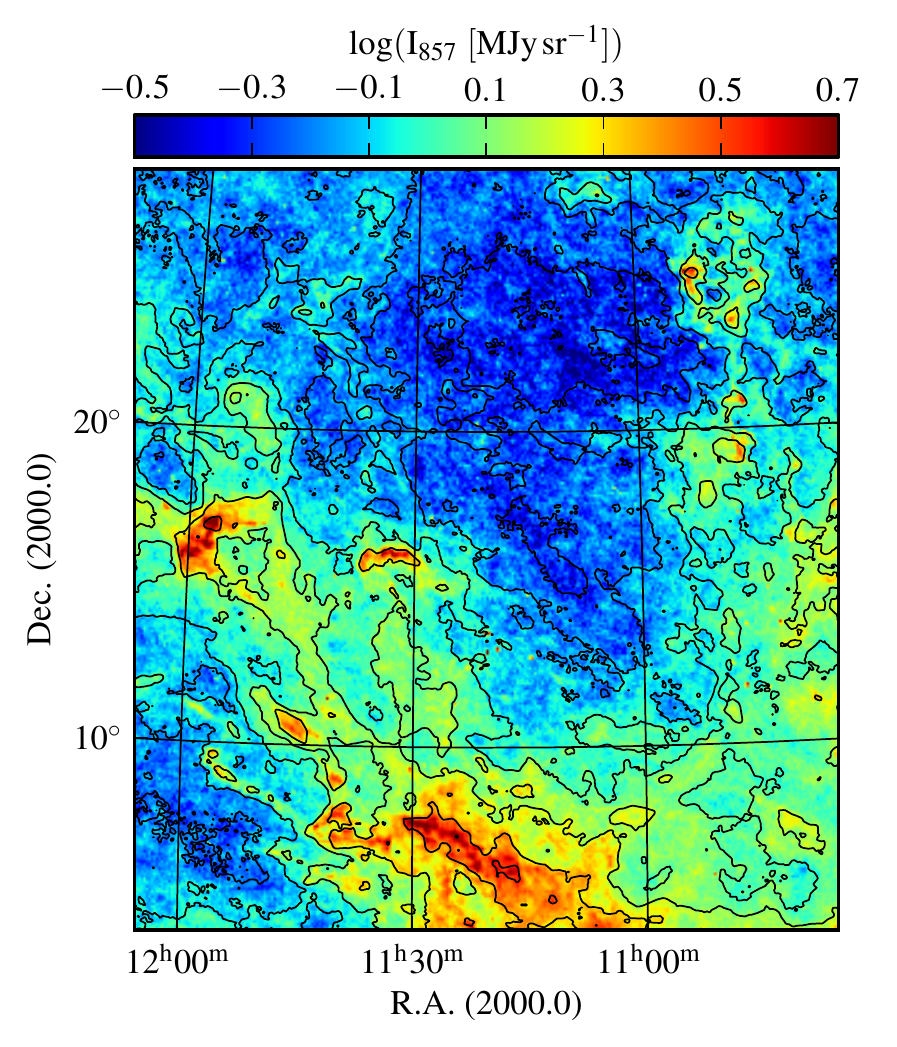}}\\
  \caption{\textit{Left}: Column density map of EBHIS integrated between $-100\,\mathrm{km}\,\mathrm{s}^{-1} \leq v_{\mathrm{LSR}} \leq +100\,\mathrm{km}\,\mathrm{s}^{-1}$. \textit{Right}: Corresponding unsmoothed Planck map at $857\,\mathrm{GHz}$. In both figures the black contours mark $N_{\ion{H}{i}}$ at the levels of the colour bar tick labels of the column density map.
  }
  \label{fig:ebhis-planck}
\end{figure*}

\begin{figure*}[!t]
  \centering
  \subfloat{\includegraphics[width=0.5\textwidth]{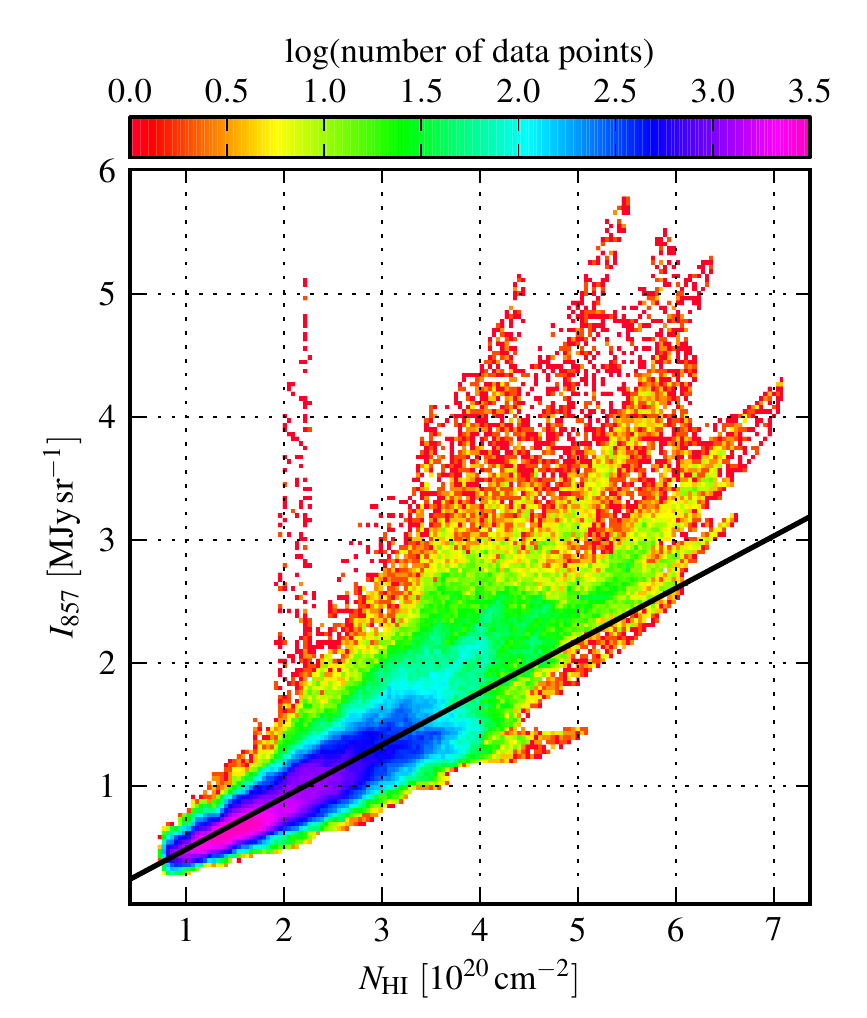}}
  \subfloat{\includegraphics[width=0.5\textwidth]{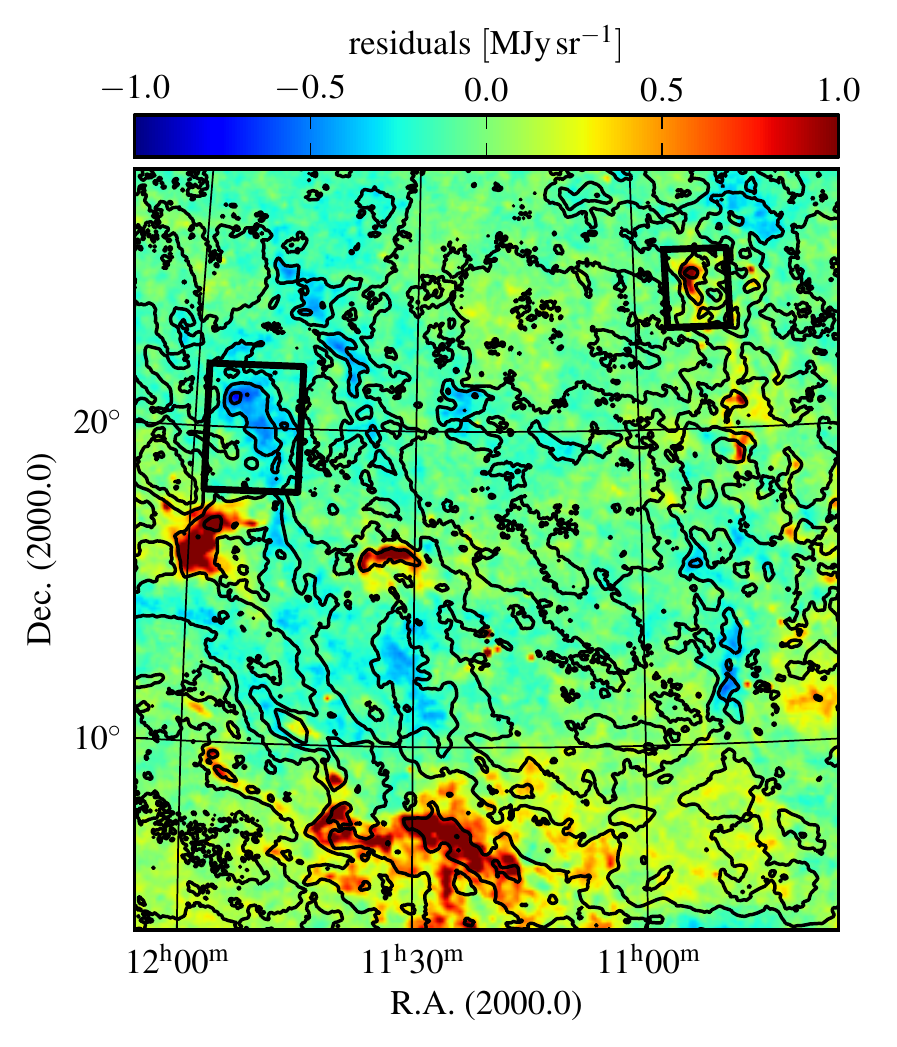}}\\
   \caption{\textit{Left}: Correlation plot $N_{\ion{H}{i}}$ vs. $857\,\mathrm{GHz}$ for the entire field. The black solid line is our best fit of Eq. \eqref{eq:1} to the data below $N_{\ion{H}{i}}=2\times10^{20}\,\mathrm{cm}^{-2}$. \textit{Right}: Residual map at $857\,\mathrm{GHz}$. The residuals are derived by subtracting for each pixel the FIR sky brightness calculated from Eq. \eqref{eq:1} using the linear-fit parameters. The black contours denote $N_{\ion{H}{i}}$ as in Fig. \ref{fig:ebhis-planck}. The black rectangles mark the locations of two $\ion{H}{i}$ clouds (IVC\,1 on the left, IVC\,2 on the right) which are presented in more detail in Sect. \ref{sec:individual-clouds}.}
   \label{fig:hi-ir857}
 \end{figure*}

\section{Analysis of the entire field}
\label{sec:entire-field}

In this section we present the analyses of the entire field: first the $\ion{H}{i}$-FIR correlation, second the modified black-body fitting and third the estimation of molecular hydrogen column densities from the FIR emission.

\subsection{Global correlation of gas and dust}
\label{sec:global-hi-ir}

An $\ion{H}{i}$ column density map of the entire studied field integrated across $-100\,\mathrm{km}\,\mathrm{s}^{-1} \leq v_{\mathrm{LSR}} \leq +100\,\mathrm{km}\,\mathrm{s}^{-1}$ and the corresponding Planck data at $857\,\mathrm{GHz}$ are shown in Fig. \ref{fig:ebhis-planck}. From these maps the close correlation between gas and dust is evident. The field is located near a region of low $N_{\ion{H}{i}}$ which is surrounded by the IV Arch and Spur. From the bottom towards the north-east there are large amounts of local gas that is part of high-latitude $\ion{H}{i}$-shells probably directly connected to Loop I and the North Polar Spur \citep{Puspitarini2012}.

We create $\ion{H}{i}$-FIR correlation plots (Fig. \ref{fig:hi-ir857}, left for $857\,\mathrm{GHz}$) and use Eq. \eqref{eq:1} to fit dust emissivities which are listed in Table \ref{tab:hi-ir} (second column). The correlation parameters are estimated for $N_{\ion{H}{i}} \leq 2.0 \times 10^{20}\,\mathrm{cm}^{-2}$ which is the empirical threshold for the $\ion{H}{i}$-H$_{2}$ transition found in other studies and also for one particular IVC within the field (Sect. \ref{sec:ivc2}).

\begin{table}[!t]
  \caption{Dust emissivities across the entire field, IVC\,1 and IVC\,2. The values are derived by fitting Eq. \eqref{eq:1} to the $\ion{H}{i}$-FIR correlation plots (Figs. \ref{fig:hi-ir857}, \ref{fig:ivcs-hi-ir}).}  
  \label{tab:hi-ir}
  \small
  \centering
  \begin{tabular}{cccc}
    \hline\hline
    $\nu$ [GHz] & $\epsilon_{\nu}^{\mathrm{total}}$ $[\frac{\mathrm{MJy}\,\mathrm{sr}^{-1}}{10^{20}\,\mathrm{cm}^{-2}}]$ & $\epsilon_{\nu}^{\mathrm{IVC\,1}}$ $[\frac{\mathrm{MJy}\,\mathrm{sr}^{-1}}{10^{20}\,\mathrm{cm}^{-2}}]$ & $\epsilon_{\nu}^{\mathrm{IVC\,2}}$ $[\frac{\mathrm{MJy}\,\mathrm{sr}^{-1}}{10^{20}\,\mathrm{cm}^{-2}}]$\\  
     \hline
     $353$ & $0.0505 \pm 0.0002$ & $0.0333 \pm 0.0004$ & $0.050 \pm 0.003$ \\
     $545$ & $0.1577 \pm 0.0002$ & $0.0789 \pm 0.0005$ & $0.185 \pm 0.005$ \\
     $857$ & $0.4243 \pm 0.0006$ & $0.239 \pm 0.001$ & $0.56 \pm 0.01$ \\
     $3000$ & $0.5597 \pm 0.0007$ & $0.631 \pm 0.001$ & $0.74 \pm 0.02$ \\
     \hline
  \end{tabular}
\end{table}

There is a large scatter around a linear relation (Fig. \ref{fig:hi-ir857}, left). Many clouds in the field have a FIR excess, but some also have a deficiency in brightness. We obtain a residual map (Fig. \ref{fig:hi-ir857}, right) by subtracting in each pixel the FIR intensity that is expected from Eq. \eqref{eq:1} using the fitted linear parameters. In particular we find two IVCs with completely different FIR emission despite their similarity in $\ion{H}{i}$. These two clouds are marked by the black boxes in Fig. \ref{fig:hi-ir857} (right). Section \ref{sec:individual-clouds} gives a detailed description of these two IVCs.

\subsection{Maps of dust temperature and optical depth}
\label{sec:map-of-dus-tem-and-opt-dep}

Our FIR data probe the thermal dust emission. We fit modified black-body spectra of the form 
\begin{equation}
\label{eq:2}
 I_{\nu}(\tau_{857},\beta,T_{\mathrm{D}}) = \tau_{857} \times \left( \frac{\nu}{\nu_{0}} \right)^{\beta} \times B_{\nu}(T_{\mathrm{D}}).
\end{equation}
Parameters are the dust optical depth $\tau_{857}$, which we choose at $857\,\mathrm{GHz}$; the dust temperature $T_{\mathrm{D}}$; and the spectral index $\beta$ of the power law emissivity $\left(\nu/\nu_{0}\right)^{\,\beta}$. We normalise the dust emission spectrum to $\nu_{0}=857\,\mathrm{GHz}$. From \citet{Planckcollaboration2011XXIV} we adopt $\beta=1.8$. Furthermore, we smooth the FIR data to the lowest resolution, which is the $353\,\mathrm{GHz}$ map. The noise values given in \citet{Planckcollaboration2013I} and \citet{MivilleDeschenes2005b} are used as weights for the modified black-body fits.

There may be offsets in the FIR data that are not related to the galactic gas distribution. These are corrected for by subtracting offsets that are estimated in \citet{Planckcollaboration2011XIX}, where reference pixels with $N_{\ion{H}{i}}\leq 2.0 \times 10^{19}\,\rm{cm}^{-2}$ are used to infer the smooth background FIR radiation. 

\begin{figure*}[!t]
 \centering
  \subfloat{\includegraphics[width=0.5\textwidth]{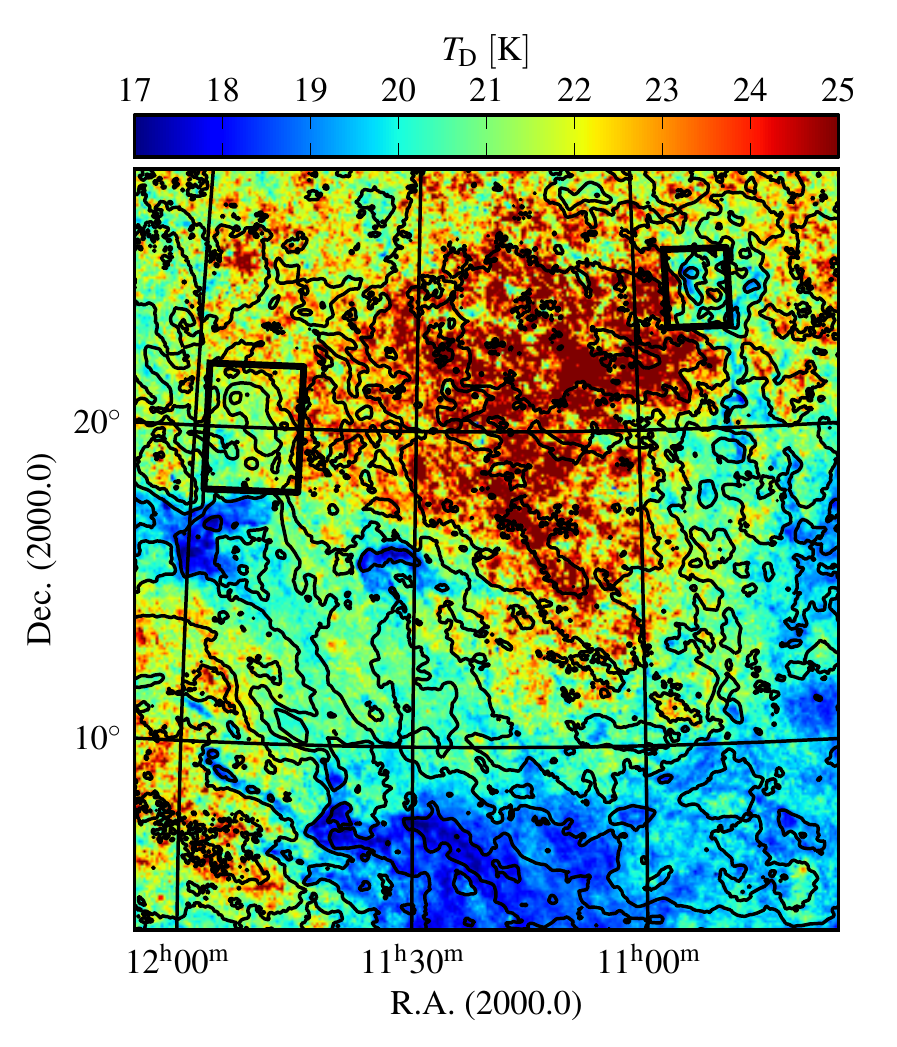}}
  \subfloat{\includegraphics[width=0.5\textwidth]{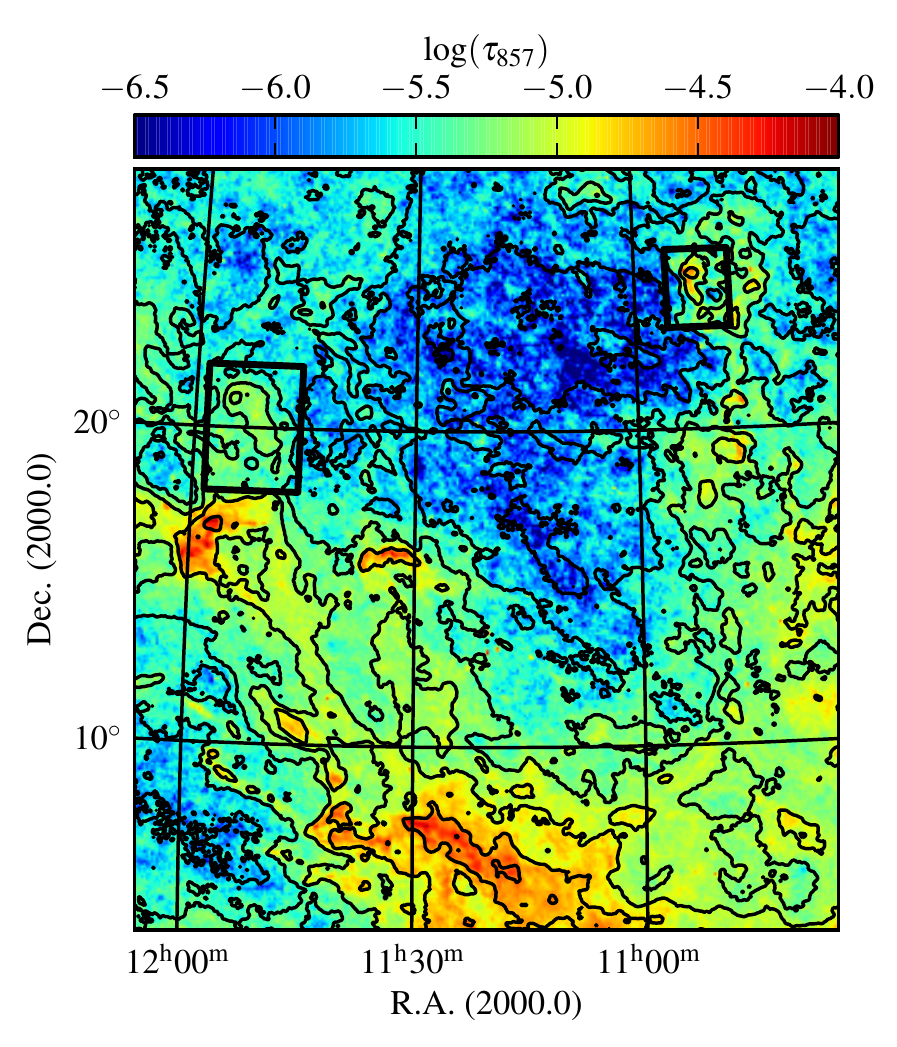}}\\
  \caption{Maps of dust temperature $T_{\mathrm{D}}$ (\textit{left}) and dust optical depth $\tau_{857}$ (\textit{right}) at the resolution of the $353\,\mathrm{GHz}$ data. Black contours mark $N_{\ion{H}{i}}$ as in Fig. \ref{fig:ebhis-planck}.}
  \label{fig:maps-dust-temp-optical-depth}
\end{figure*}

By fitting Eq. \eqref{eq:2} in each pixel independently, we derive maps of dust temperature $T_{\mathrm{D}}$ and dust optical depth $\tau_{857}$ (Fig. \ref{fig:maps-dust-temp-optical-depth}). The dust temperatures have a median value of $21.2\,\mathrm{K}$. The dust optical depths $\tau_{857}$ cover about two orders of magnitude. Both $T_{\mathrm{D}}$ and $\tau_{857}$ correlate to gaseous structures. 

\begin{figure*}[!t]
 \centering
  \subfloat{\includegraphics[width=0.333\textwidth]{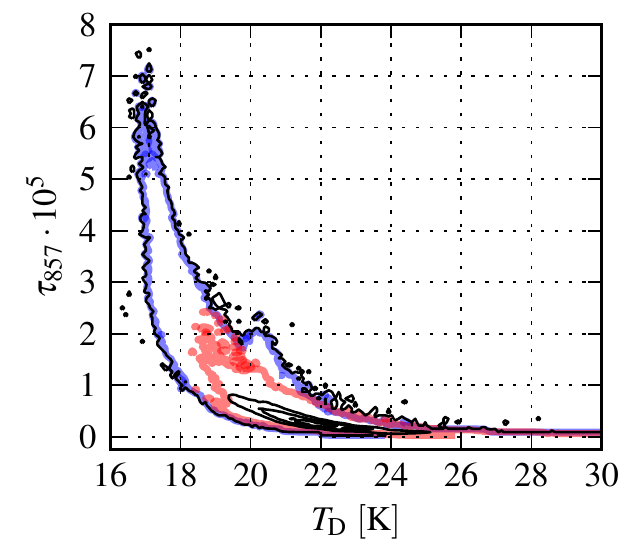}}
  \subfloat{\includegraphics[width=0.333\textwidth]{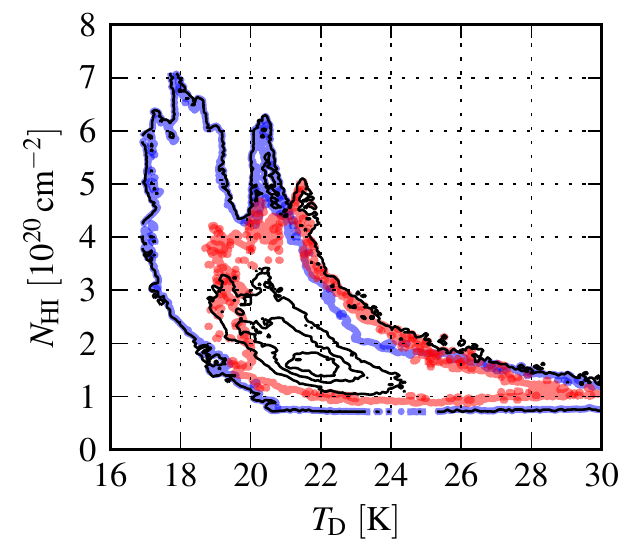}}
  \subfloat{\includegraphics[width=0.333\textwidth]{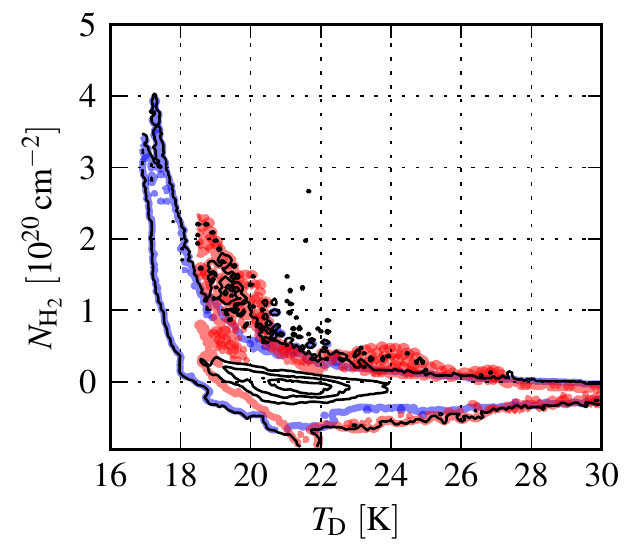}}\\
  \caption{Dependence of dust optical depth $\tau_{857}$ (\textit{left}), $\ion{H}{i}$ column density $N_{\ion{H}{i}}$ (\textit{middle}), and H$_{2}$ column density $N_{\mathrm{H}_{2}}$ (\textit{right}) on dust temperature $T_{\mathrm{D}}$. The $T_{\mathrm{D}}-\tau_{857}$ plot is done for FIR data at the resolution of the $353\,\mathrm{GHz}$ data while the other two plots are for FIR data smoothed to the EBHIS resolution. The black contours show the number density distribution of all data points. The blue contour traces the pixels that contain more $\ion{H}{i}$ between $-20\,\mathrm{km}\,\mathrm{s}^{-1} \leq v_{\mathrm{LSR}} \leq +20\,\mathrm{km}\,\mathrm{s}^{-1}$ than in the IVC range $-60\,\mathrm{km}\,\mathrm{s}^{-1} \leq v_{\mathrm{LSR}} \leq -20\,\mathrm{km}\,\mathrm{s}^{-1}$, whereas the red contour traces pixels that have more $\ion{H}{i}$ in the IVC regime. In Sect. \ref{sec:map-of-mol-hyd-col-den} we describe how we derive an H$_{2}$ map from the FIR excess emission. Negative $N_{\mathrm{H}_{2}}$ 
are a result of the derivation.}
  \label{fig:dust-temp-opticaldepth}
\end{figure*}

In Fig. \ref{fig:dust-temp-opticaldepth} (left, middle) we compare dust temperatures to dust optical depths and $\ion{H}{i}$ column densities. In general, low dust temperatures are associated with large optical depths and vice versa (Fig. \ref{fig:dust-temp-opticaldepth}, left). A similar correlation exists between $T_{\mathrm{D}}$ and $N_{\mathrm{\ion{H}{i}}}$ (Fig. \ref{fig:dust-temp-opticaldepth}, middle). The IVC gas appears to contain warmer dust grains compared to local structures confirming previous studies \citep[e.g.][]{Planckcollaboration2011XXIV}. This is related to the galactic fountain origin of IVCs by which dust grains are shattered and the number of very small grains is increased (see e.g. \citealt{Jones2011} for a discussion of the destruction and survival of dust grains in the ISM in general).

We note that the fitted dust temperatures do not reflect the actual dust temperatures \citep{Planckcollaboration2011XXIV}. A simple modified black body is an average over the line of sight and its parameters are biased towards regions of bright emission. Nevertheless, spatial variations of this fitted dust temperature are related to changes in the spectral energy distribution.

\subsection{Estimation of the molecular hydrogen column density}
\label{sec:map-of-mol-hyd-col-den}

The thermal dust emission acts as a tracer of the total gas column density. The linear $\ion{H}{i}$-FIR correlation covers the range where hydrogen is neutral, whereas the FIR excess emission traces H$_{2}$. Thus, maps of molecular hydrogen column density can be calculated from the total FIR emission \citep[e.g.][]{Dame2001}.

\subsubsection{FIR emission and molecular hydrogen}
\label{sec:fir-h2}

We assume that the excess emission is solely due to H$_{2}$. Any influences due to H$^{+}$ are considered to be negligible and changes of the dust-to-gas ratio or emissivity properties of the grains are omitted. The simple $\ion{H}{i}$-FIR correlation in Eq. \eqref{eq:1} is generalised to account for H$_{2}$:
\begin{equation}
  I_{\nu} = R_{\nu} + \epsilon_{\nu} \times N_{\mathrm{H}} = R_{\nu} + \epsilon_{\nu} \times (N_{\ion{H}{i}} + 2 N_{\mathrm{H}_{2}}).  
\end{equation}
Solving for $N_{\mathrm{H}_{2}}$ yields
\begin{equation}
\label{eq:nh2-map}
  N_{\mathrm{H}_{2}} = \frac{1}{2} \left( \frac{I_{\nu}-R_{\nu}}{\epsilon_{\nu}}-N_{\ion{H}{i}} \right).
\end{equation}
In each pixel of the map we calculate $N_{\mathrm{H}_{2}}$ from the total $\ion{H}{i}$ column density and the $857\,\mathrm{GHz}$ data smoothed to the EBHIS resolution of $10.8\,\arcmin$. For the correlation parameters, we use the values derived for the entire field (Table \ref{tab:hi-ir}, first column).

A map of $N_{\mathrm{H}_{2}}$ is shown in Fig. \ref{fig:map-nh2}; $\ion{H}{i}$ structures with local velocities are prominent, with $N_{\mathrm{H}_{2}}$ in the range $1-4\times 10^{20}\,\mathrm{cm}^{-2}$. Negative H$_{2}$ column densities result from the derivation and are not physically meaningful. We do not clip these values in order to show the uncertainties of the H$_{2}$ map.

In Fig. \ref{fig:dust-temp-opticaldepth} (right) the relation between $T_{\mathrm{D}}$ and $N_{\mathrm{H}_{2}}$ is shown. For $T_{\mathrm{D}}\gtrsim22\,\mathrm{K}$ no significant amount of H$_{2}$ is observed, neither in local nor IVC gas. 

The correlation parameters $R_{\nu}$ and $\epsilon_{\nu}$ that are applied in Eq. \eqref{eq:nh2-map} are statistically well constrained. In order to estimate an uncertainty for our $N_{\mathrm{H}_{2}}$ map, we use the scatter in the residuals from the $\ion{H}{i}$-FIR correlation after subtracting the best fit. For $857\,\mathrm{GHz}$ this results in a $1\sigma$ scatter of $0.2\times10^{20}\,\mathrm{cm}^{-2}$ for $N_{\mathrm{H}_{2}}$. In addition to this statistical error, systematic errors contribute, such as local changes in the grain emissivity, in metallicity, or in dust-to-gas ratio, all of which may mimic the presence of H$_{2}$.
 
The molecular gas we infer in the field is not observed in the large-scale CO survey on the northern high-latitude sky \citep{Hartmann1998}. Furthermore, the all-sky maps of CO emission extracted from the Planck foreground modelling also do not contain any CO in this area of the sky \citep{Planckcollaboration2013XIII}.

\subsubsection{Validation of the $N_{\mathrm{H}_{2}}$ map}
\label{sec:validation-nh2-map}

\begin{figure}[!t]
 \centering
  \resizebox{\hsize}{!}{\includegraphics[width=1.\textwidth]{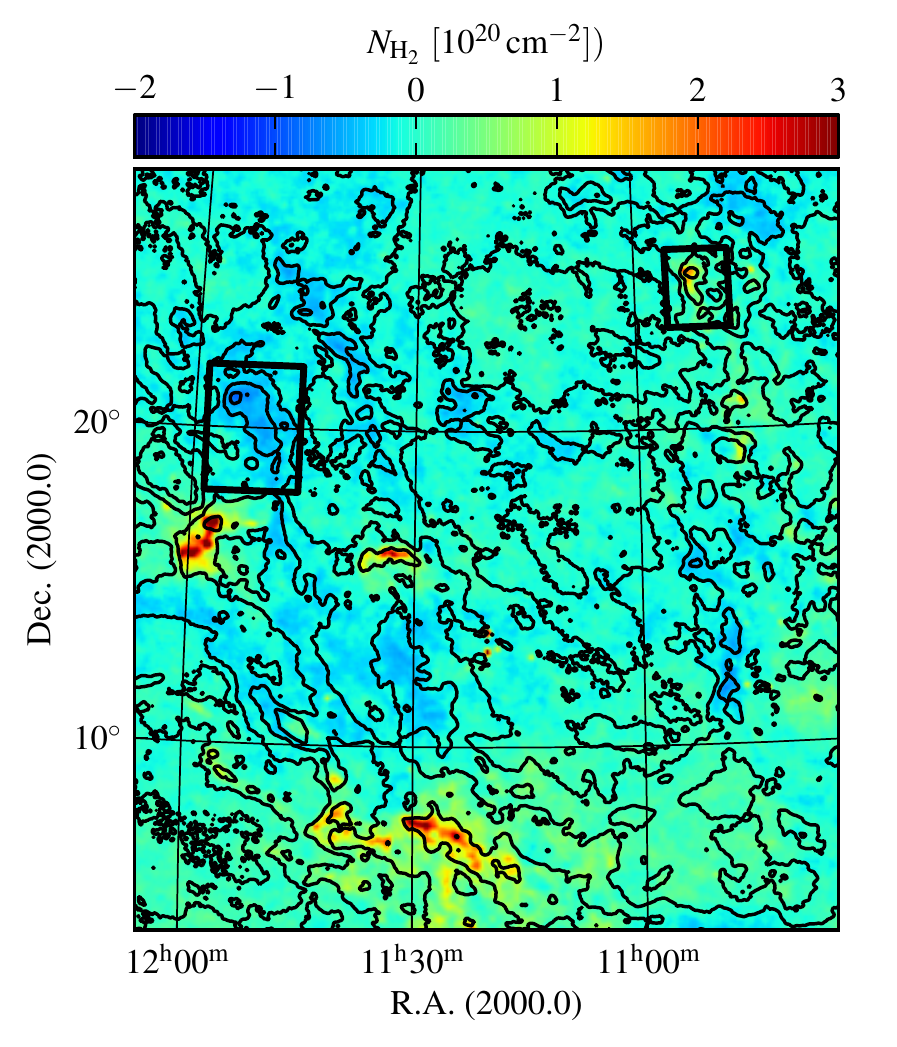}}
  \caption{Map of molecular hydrogen column density $N_{\mathrm{H}_{2}}$ with a $1-\sigma$ uncertainty of $0.2\times10^{20}\,\mathrm{cm}^{-2}$. Negative values result from the derivation and are not physically meaningful. Black contours mark $N_{\ion{H}{i}}$ as in Fig. \ref{fig:ebhis-planck}.}
  \label{fig:map-nh2}
\end{figure}

The $N_{\mathrm{H}_{2}}$ map in Fig. \ref{fig:map-nh2} is derived directly from the $857\,\mathrm{GHz}$ intensity and the $\ion{H}{i}$ column density using Eq. \eqref{eq:nh2-map}. Together, $\ion{H}{i}$ and H$_{2}$ approximate the distribution of the total hydrogen column density $N_{\mathrm{H}}$. The dust optical depths $\tau_{857}$ also trace $N_{\mathrm{H}}$, in a manner completely independent from Eq. \eqref{eq:nh2-map}, since $\tau_{857}$ is obtained from modified black-body fits (Eq. \ref{eq:2}) to the FIR spectra.

\begin{figure}[!t]
 \centering
  \resizebox{\hsize}{!}{\includegraphics[width=1.\textwidth]{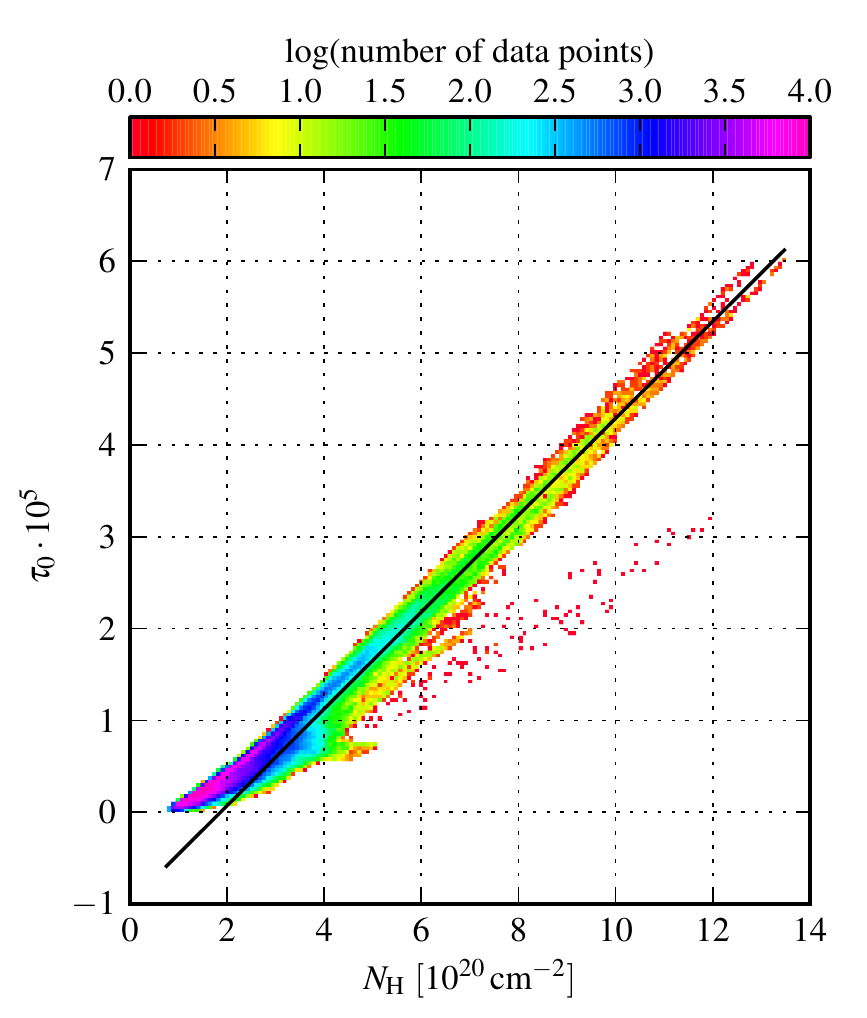}}
  \caption{Comparison between the estimated total hydrogen column densities $N_{\mathrm{H}}$ using Eq. \eqref{eq:nh2-map} and the fitted dust optical depth $\tau_{857}$ from modified black-body fits (Eq. \ref{eq:2}). For $N_{\mathrm{H}}\geq 5\times 10^{20}\,\mathrm{cm}^{-2}$ a linear relation is fitted and shown by the black line.}
  \label{fig:tau-nh}
\end{figure}

For the validation of our H$_{2}$ map, we compare the total hydrogen column density $N_{\mathrm{H}}=N_{\ion{H}{i}}+2N_{\mathrm{H}_{2}}$ to the dust optical depth $\tau_{857}$. For this, negative $N_{\mathrm{H}_{2}}$ values are clipped. In Fig. \ref{fig:tau-nh} we plot $N_{\mathrm{H}}$ against $\tau_{857}$ for each pixel of the entire field. Above $N_{\mathrm{H}}=5\times10^{20}\,\mathrm{cm}^{-2}$ we fit a linear function to the data, yielding the slope $\tau_{857} / N_{\mathrm{H}}\simeq5.1\times10^{-26}\,\mathrm{cm}^{-2}$. This agrees well with the value of $5.1\times10^{-26}\,\mathrm{cm}^{-2}$ for $857\,\mathrm{GHz}$ given by \citet{Boulanger1996}. Varying the fitting threshold in the range $4-6\times 10^{20}\,\mathrm{cm}^{-2}$ results in values of $\tau_{857} / N_{\mathrm{H}}$ which range between $5.0-5.3\times10^{-26}\,\mathrm{cm}^{-2}$. Thus Eq. \eqref{eq:nh2-map} gives a good approximation of the molecular hydrogen column densities from the FIR intensity and the $\ion{H}{i}$ distribution directly.

Figure \ref{fig:tau-nh} shows a significant enhancement of $\tau_{857}$ with respect to the linear fit below $N_{\mathrm{H}}\simeq2\times10^{20}\,\mathrm{cm}^{-2}$. This deviation is probably related to an increasing dust contribution from ionised hydrogen which is not considered here. In principle, a map of H$^{+}$ can be derived from the $N_{\mathrm{H}}-\tau_{857}$ correlation by shifting the points above the linear fit towards higher $N_{\mathrm{H}}$.

\section{Analysis of individual clouds}
\label{sec:individual-clouds}

In the following we focus on two IVCs distinguished by their different FIR properties, as is evident from the global residuals in Fig. \ref{fig:hi-ir857} (right), where the locations of the clouds are given by the black rectangles. The FIR-dim IVC at the top-left of the field (IVC\,1) belongs to the IV Spur, the FIR-bright cloud (IVC\,2) to the IV Arch. 

\begin{figure*}[!t]
  \centering
  \subfloat{\includegraphics[width=0.5\textwidth]{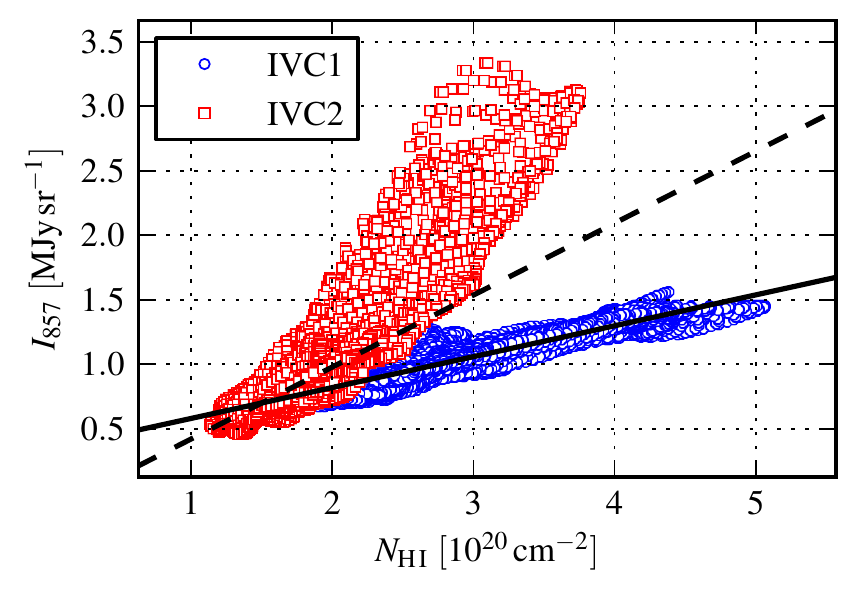}}
  \subfloat{\includegraphics[width=0.5\textwidth]{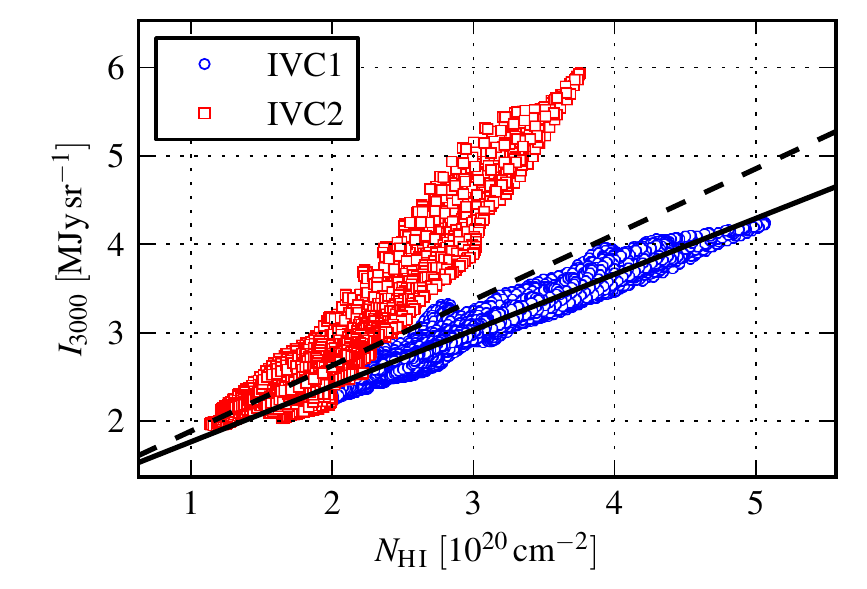}}\\
  \caption{$\ion{H}{i}$-FIR correlations for IVC\,1 and IVC\,2 at $857\,\mathrm{GHz}$ (\textit{left}) and $3000\,\mathrm{GHz}$ (\textit{right}). The $\ion{H}{i}$ is integrated between $-100\,\mathrm{km}\,\mathrm{s}^{-1} \leq v_{\mathrm{LSR}} \leq +100\,\mathrm{km}\,\mathrm{s}^{-1}$. The black solid line is the linear fit applying Eq. \eqref{eq:1} for IVC\,1 and the dashed black line for IVC\,2. For IVC\,2 the fit is restricted to $N_{\ion{H}{i}} \leq 2.0 \times 10^{20}\,\mathrm{cm}^{-2}$, above which a FIR excess due to H$_{2}$ is evident. For IVC\,2 the spread in $N_{\ion{H}{i}}$ for given $I_{\nu}$ is due to a second FIR maximum that is not associated with an $\ion{H}{i}$ peak (Fig. \ref{fig:ivc2-sub}).
  }
  \label{fig:ivcs-hi-ir}
\end{figure*}

\subsection{IVC\,1}
\label{sec:ivc1}
 
The cloud IVC\,1 is located at (R.A., Dec) $=$ ($11^{\mathrm{h}} 52^{\mathrm{m}}$, $20^{\circ}30\arcmin$) and it is FIR dim. \citet{Desert1988} mention this region as a FIR-deficient IVC at $-30\,\mathrm{km}\,\mathrm{\mathrm{s}}^{-1}$, although the cloud structure cannot be resolved by their angular resolution of $\simeq 40\arcmin$ in $\ion{H}{i}$.
 
In Fig. \ref{fig:ivcs-hi-ir} the $\ion{H}{i}$-FIR correlations are shown for $857\,\mathrm{GHz}$ and $3000\,\mathrm{GHz}$ and for $\ion{H}{i}$ integrated between $-100\,\mathrm{km}\,\mathrm{s}^{-1} \leq v_{\mathrm{LSR}} \leq 100\,\mathrm{km}\,\mathrm{\mathrm{s}}^{-1}$. For IVC\,1 we fit Eq. \eqref{eq:1} to the entire data range, since the correlation is linear throughout. The resulting dust emissivities $\epsilon_{\nu}^{\mathrm{IVC\,1}}$ are compiled in Table \ref{tab:hi-ir} (third column). In the Planck frequency range the emissivities are lower than our estimated values over the entire field. There is no indication of FIR excess emission such as would indicate significant amounts of molecular hydrogen. In the CO survey performed by \citet{Hartmann1998}, IVC\,1 is not detected.
 
\begin{figure}[!t]
  \centering
  \resizebox{\hsize}{!}{\includegraphics{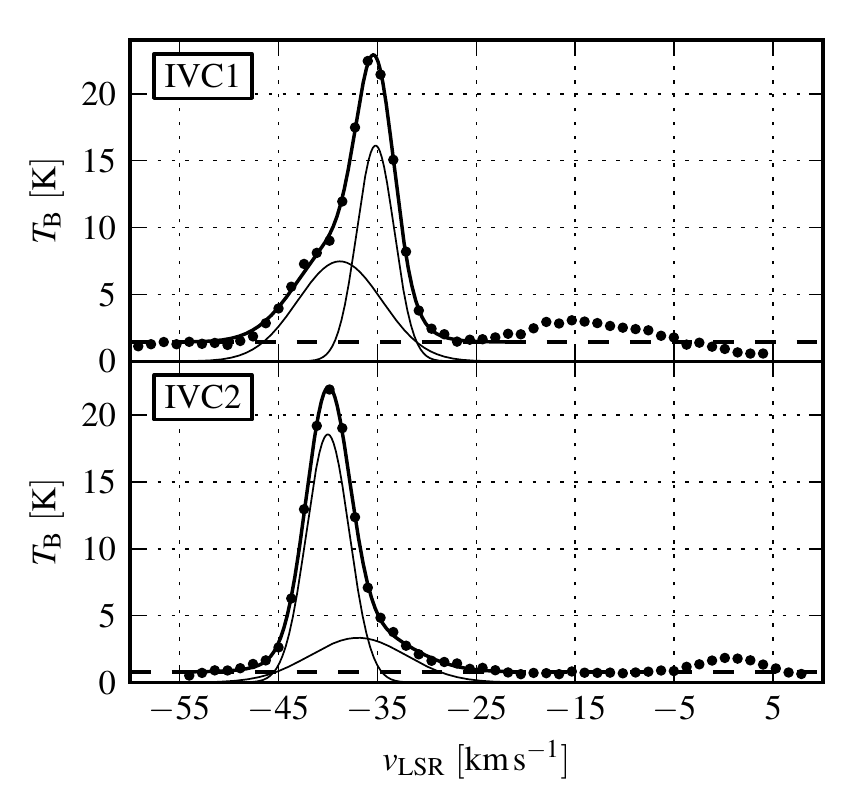}}
  \caption{$\ion{H}{i}$ spectra of IVC\,1 (\textit{top}) and IVC\,2 (\textit{bottom}) at their maximum brightness temperature. Both spectra are fitted by two Gaussians plus a constant offset.}
  \label{fig:ivcs-spectra-fit}
\end{figure}

The $\ion{H}{i}$ spectrum at the peak brightness temperature is shown in Fig. \ref{fig:ivcs-spectra-fit} (top). It is a two-component spectral profile consisting of a bright line with $\mathrm{FWHM}\simeq4.5\,\mathrm{km}\,\mathrm{\mathrm{s}}^{-1}$ centred on $-35.2\,\mathrm{km}\,\mathrm{\mathrm{s}}^{-1}$ and a second fainter one at $-38.8\,\mathrm{km}\,\mathrm{\mathrm{s}}^{-1}$ with $\mathrm{FWHM}\simeq10.0\,\mathrm{km}\,\mathrm{\mathrm{s}}^{-1}$. About $50\%$ of the $\ion{H}{i}$ is in the colder phase. In the spectrum there is not much $\ion{H}{i}$ at other radial velocities. A column density map of IVC\,1 integrated over the spectral range between $-60\,\mathrm{km}\,\mathrm{s}^{-1} \leq v_{\mathrm{LSR}} \leq -20\,\mathrm{km}\,\mathrm{\mathrm{s}}^{-1}$ is presented in Fig. \ref{fig:ivc1-int}. This shows a peak value of $N_{\ion{H}{i}} \simeq 3.9 \times 10^{20}\,\mathrm{cm}^{-2}$. The cloud consists of several cold clumps that are well connected in $\ion{H}{i}$. Its morphology is curved and filamentary.

The dust spectrum of IVC\,1 at its maximum $N_{\ion{H}{i}}$ is shown in Fig. \ref{fig:dust-spectra} (top). The emission is well approximated by a modified black body with $\beta=1.8$. The fitted dust parameters are given in Table \ref{tab:phys-prop-clou}.

\subsection{IVC\,2}
\label{sec:ivc2}

\subsubsection{$\ion{H}{i}$ and FIR data}
\label{sec:ivc2-hi-fir}
 
The cloud IVC\,2 is located at the top right of the field at (R.A., Dec) $=$ ($10^{\mathrm{h}}52^{\mathrm{m}}$, $25^{\circ}$). The cloud is selected because of its bright FIR emission and its similarity to IVC\,1 in $\ion{H}{i}$.

In Fig. \ref{fig:ivcs-hi-ir} the $\ion{H}{i}$-FIR correlation plots are presented. The linear fits are performed below $N_{\ion{H}{i}} \leq 2.0 \times 10^{20}\,\mathrm{cm}^{-2}$ where a linear model is thought to be valid. This is the threshold we also apply for the entire field (Sect. \ref{sec:global-hi-ir}). At the lowest $N_{\ion{H}{i}}$, the cloud's FIR emission is equal to that of IVC1 (Fig. \ref{fig:ivcs-hi-ir}), although the fitted FIR emissivities of IVC\,2 are notably larger (Table \ref{tab:hi-ir}, fourth column). The emissivity increases even more when larger column densities are considered. 

The $\ion{H}{i}$ spectrum at the highest brightness temperature (Fig. \ref{fig:ivcs-spectra-fit}, bottom) resembles the spectrum of IVC\,1. It is well approximated by two Gaussians with $\mathrm{FWHM}\simeq5.1\,\mathrm{km}\,\mathrm{\mathrm{s}}^{-1}$ centred on $-40.0\,\mathrm{km}\,\mathrm{\mathrm{s}}^{-1}$ and $\mathrm{FWHM}\simeq11.5\,\mathrm{km}\,\mathrm{\mathrm{s}}^{-1}$ at $-37.0\,\mathrm{km}\,\mathrm{\mathrm{s}}^{-1}$. In IVC\,2 about $70\%$ of the $\ion{H}{i}$ is in the colder phase. As with IVC\,1, there is not much $\ion{H}{i}$ at different radial velocities in the spectrum. Figure \ref{fig:ivc2-int} shows a column density map integrated between $-60\,\mathrm{km}\,\mathrm{s}^{-1} \leq v_{\mathrm{LSR}} \leq -20\,\mathrm{km}\,\mathrm{\mathrm{s}}^{-1}$ with a maximum of $N_{\ion{H}{i}} \simeq 3.1 \times 10^{20}\,\mathrm{cm}^{-2}$, which is less than for IVC\,1. Structures emerge from the central core in several directions showing other cold clumps.

The dust spectrum of IVC\,2 at its maximum $N_{\ion{H}{i}}$ is shown in Fig. \ref{fig:dust-spectra} (bottom). As for IVC\,1, the emission is well described by a modified black body with $\beta=1.8$. The fitted dust parameters are given in Table \ref{tab:phys-prop-clou}. The dust optical depth in IVC\,2 is about four times larger, and the fitted grain temperatures are slightly lower, than in IVC\,1.

The angular resolution of the FIR data is higher than in the $\ion{H}{i}$ (Table \ref{tab:data}). Indeed IVC\,2 shows more structure in the FIR than in $\ion{H}{i}$: east of the central core a second maximum is evident in the FIR only (Fig. \ref{fig:ivc2-sub}). In the $\ion{H}{i}$-FIR correlation of IVC\,2 (Fig. \ref{fig:ivcs-hi-ir}), this second FIR peak is causing the large spread in $N_{\ion{H}{i}}$ for given $I_{\nu}$. This second FIR peak is fitted with $(\tau_{857}$, $T_{\mathrm{D}})=((5.4 \pm 0.1) \times 10^{-5}$, $17.7 \pm 0.1 \,\mathrm{K})$, thus it has an even larger dust optical depth and lower dust temperature.

\begin{figure}[!t]
  \centering
  \resizebox{\hsize}{!}{\includegraphics{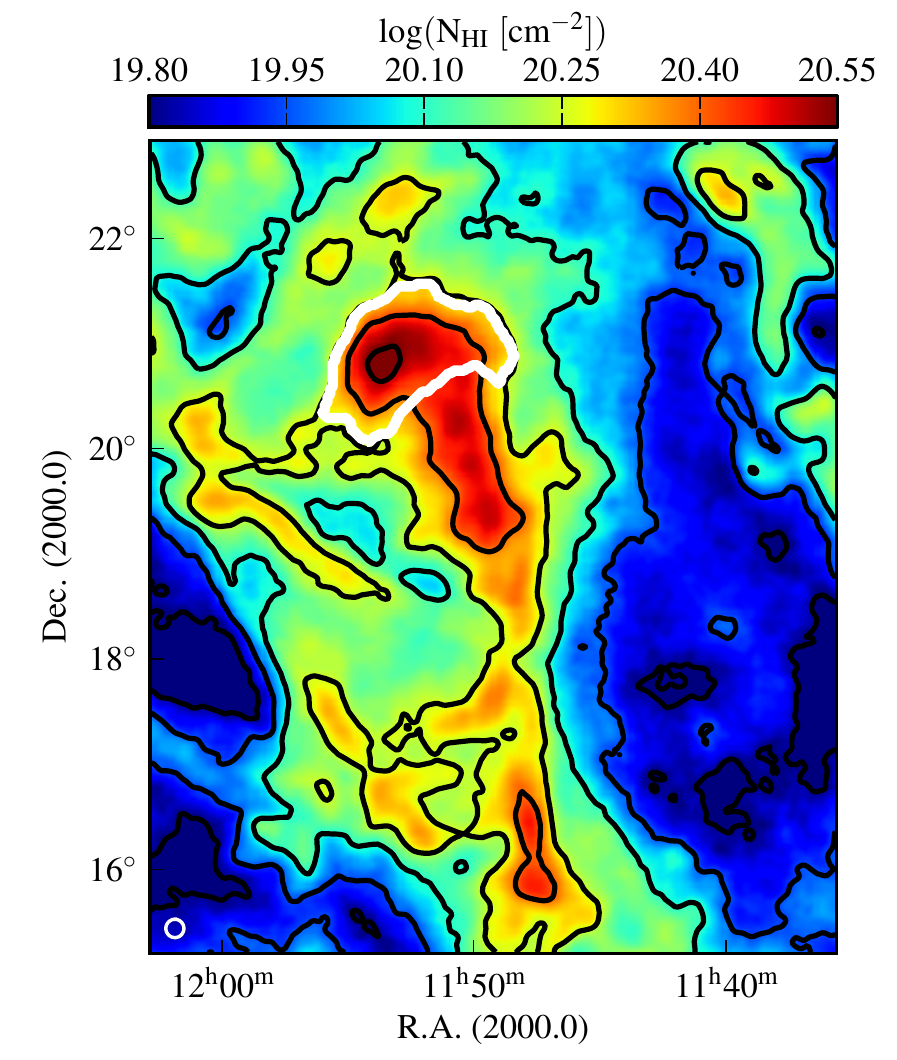}}
  \caption{Column density map of IVC\,1 integrated between $-60\,\mathrm{km}\,\mathrm{s}^{-1} \leq v_{\mathrm{LSR}} \leq -20\,\mathrm{km}\,\mathrm{\mathrm{s}}^{-1}$. The black contours mark $N_{\ion{H}{i}}$ at the levels of the colour bar tick labels. The white contour encloses the core of the IVC which is analysed in further detail in Sect. \ref{sec:estimation-of-cloud-parameters}. The white circle at the bottom left gives the angular resolution of EBHIS.}
  \label{fig:ivc1-int}
\end{figure}

\begin{figure}[!t]
 \centering
  \resizebox{\hsize}{!}{\includegraphics[width=1.\textwidth]{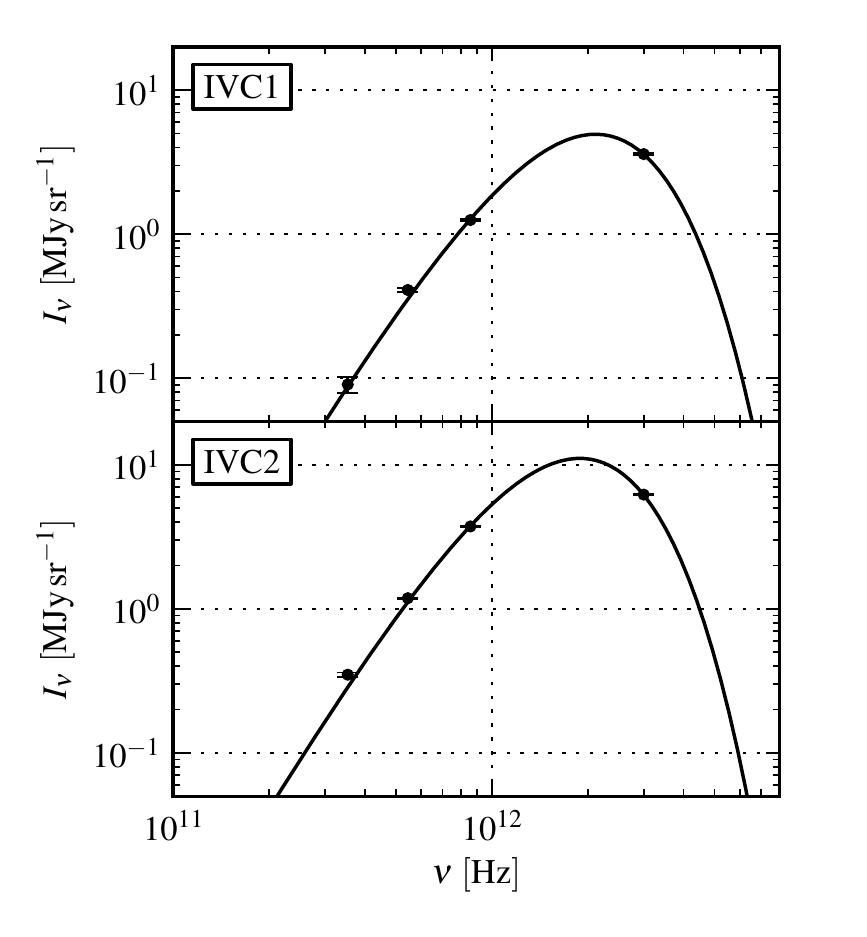}}
  \caption{FIR dust spectra and fitted modified black bodies with $\beta=1.8$. The spectra are taken at the position of the  largest $N_{\ion{H}{i}}$ for IVC\,1 (\textit{top}) and IVC\,2 (\textit{bottom}). The fitted dust parameters are $(\tau_{857}$, $T_{\mathrm{D}})_{\mathrm{IVC\,1}}=((8.3 \pm 0.4) \times 10^{-6}$, $21.2 \pm 0.2 \,\mathrm{K})$ and $(\tau_{857}$, $T_{\mathrm{D}})_{\mathrm{IVC\,2}}=((3.2 \pm 0.1) \times 10^{-5}$, $19.1 \pm 0.2 \,\mathrm{K})$.}
  \label{fig:dust-spectra}
\end{figure}

\subsubsection{H$_{2}$ and CO data}
\label{sec:ivc2-h2-co}

\citet{Desert1990a} report on pointed $^{12}$CO observations of this particular cloud. The pointing positions are indicated in Fig. \ref{fig:ivc2-sub} by the black diamonds. They detect the central core in $^{12}$CO($1 \rightarrow 0$) and the eastern FIR-peak in $^{12}$CO($2 \rightarrow 1$). The CO emission is at $v_{\mathrm{LSR}}\simeq-40\,\mathrm{km}\,\mathrm{s}^{-1}$ which is consistent with the radial velocity of the $\ion{H}{i}$ (Fig. \ref{fig:ivcs-spectra-fit}, bottom). \citeauthor{Desert1990a} report on highly varying line intensities within two beams which indicates small-scale structure in the cloud. From CO they infer $N_{\mathrm{H}_{2}}=10^{20}-10^{21}\,\mathrm{cm}^{-2}$. \citeauthor{Desert1990a} conclude that IVC\,2 is a molecular cloud at high radial velocity. 

Our derived  H$_{2}$ column densities (Sect. \ref{sec:map-of-mol-hyd-col-den}) for IVC\,2 are in the range $1-2\times 10^{20}\,\mathrm{cm}^{-2}$. Its largest molecular content is found at the eastern peak with $N_{\mathrm{H}_{2}}=2.3\times 10^{20}\,\mathrm{cm}^{-2}$. We estimate H$_{2}$ column densities from data with an angular resolution of $10.8\arcmin$. The CO spectra of \citeauthor{Desert1990a} were obtained with a telescope beam of $60\arcsec$ and $14\arcsec$ which cannot be compared directly to our results since they probe different spatial scales. \citeauthor{Desert1990a} use a conversion factor of $X_{\mathrm{CO}}=2.5\times 10^{20}\,\mathrm{cm}^{-2}\,(\mathrm{K}\,\mathrm{km}\,\mathrm{s}^{-1})^{-1}$ which is inherently a source of uncertainty for an individual cloud of sub-solar metallicity. Nevertheless, we can state that their estimates are compatible with what we find for IVC\,2.

The molecular fraction $f_{\mathrm{H}_{2}}$ is calculated by
\begin{equation}
\label{eq:4}
 f_{\mathrm{H}_{2}}=\frac{2N_{\mathrm{H}_{2}}}{N_{\ion{H}{i}}+2N_{\mathrm{H}_{2}}}=\frac{2N_{\mathrm{H}_{2}}}{N_{\mathrm{H}}}.
\end{equation}
By summing up $N_{\ion{H}{i}}$ and $N_{\mathrm{H}_{2}}$ in each pixel of the cloud's core, the total molecular fraction obtained is $f_{\mathrm{H}_{2}}^{\mathrm{IVC\,2}}\simeq0.4$.

With the estimated H$_{2}$ column densities we analyse the shielding conditions for H$_{2}$ and CO. \citet{Lee1996} calculate the corresponding shielding efficiencies. In their Table 10 they give the H$_{2}$ self-shielding efficiencies $\theta(N_{\mathrm{H}_{2}})$ for different H$_{2}$ column densities. For $N_{\mathrm{H}_{2}}\simeq2\times10^{20}\,\mathrm{cm}^{-2}$ they find $\theta(N_{\mathrm{H}_{2}})\simeq 7\times10^{-6}$. Thus the molecular hydrogen in IVC\,2 is efficiently self-shielding.

For CO formation one has to consider CO self-shielding, H$_{2}$, and dust shielding \citep{Lee1996}. \citeauthor{Lee1996} estimate the shielding efficiencies for these three mechanisms (their Table 11). Taking the correspondence of $N_{\mathrm{H}_{2}}$ to $N_{\mathrm{CO}}$ from \citet[][their Fig. 14]{Pineda2010} which they derive for the Taurus molecular cloud, we assume that for IVC\,2 $N_{\mathrm{CO}}\leq10^{14}\,\mathrm{cm}^{-2}$. We adopt $N_{\mathrm{CO}}=10^{14}\,\mathrm{cm}^{-2}$, $N_{\mathrm{H}_{2}}=2\times10^{20}\,\mathrm{cm}^{-2}$, and $N_{\mathrm{H}}=7.5\times10^{20}\,\mathrm{cm}^{-2}$ which gives $A_{\mathrm{v}}\simeq0.4$ \citep{Predehl1995}. The total shielding efficiency for IVC\,2 concerning CO is the product of the three contributing shielding mechanisms: $\theta_{1}(\mathrm{CO})\theta_{2}(N_{\mathrm{H}_{2}})\theta_3(A_{\mathrm{v}})\simeq0.9\times0.7\times0.1\simeq0.1$. Thus, the CO molecules are well shielded by dust mostly and CO is expected to be found within IVC\,2.

\begin{figure}[!t]
  \centering
  \resizebox{\hsize}{!}{\includegraphics{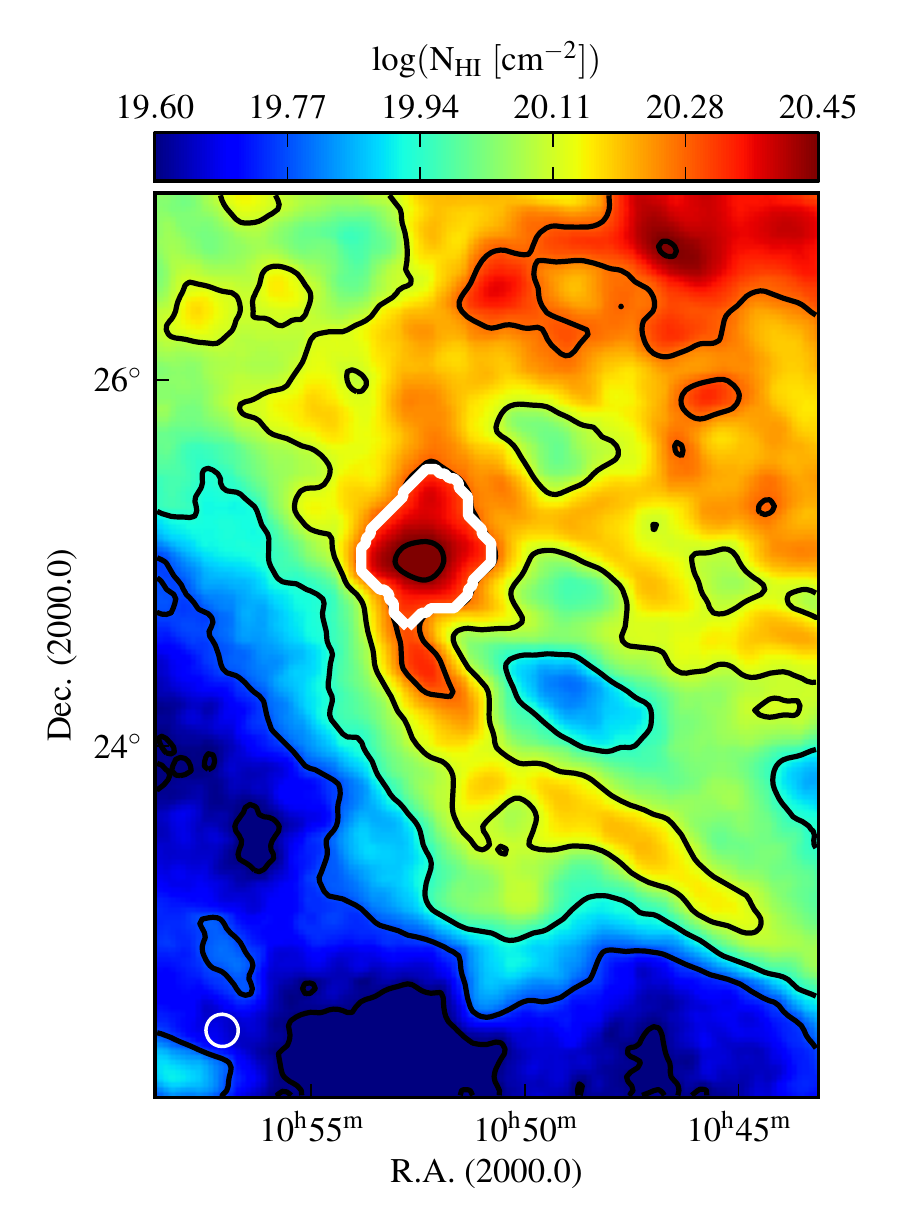}}
  \caption{Column density map of IVC\,2 integrated between $-60\,\mathrm{km}\,\mathrm{s}^{-1} \leq v_{\mathrm{LSR}} \leq -20\,\mathrm{km}\,\mathrm{\mathrm{s}}^{-1}$. The black contours mark $N_{\ion{H}{i}}$ at the levels of the colour bar tick labels. The white contour encloses the core of the IVC which is analysed in further detail in Sect. \ref{sec:estimation-of-cloud-parameters}. The white circle at the bottom left gives the angular resolution of EBHIS.}
  \label{fig:ivc2-int}
\end{figure}

\begin{figure}[!t]
  \centering
  \resizebox{\hsize}{!}{\includegraphics{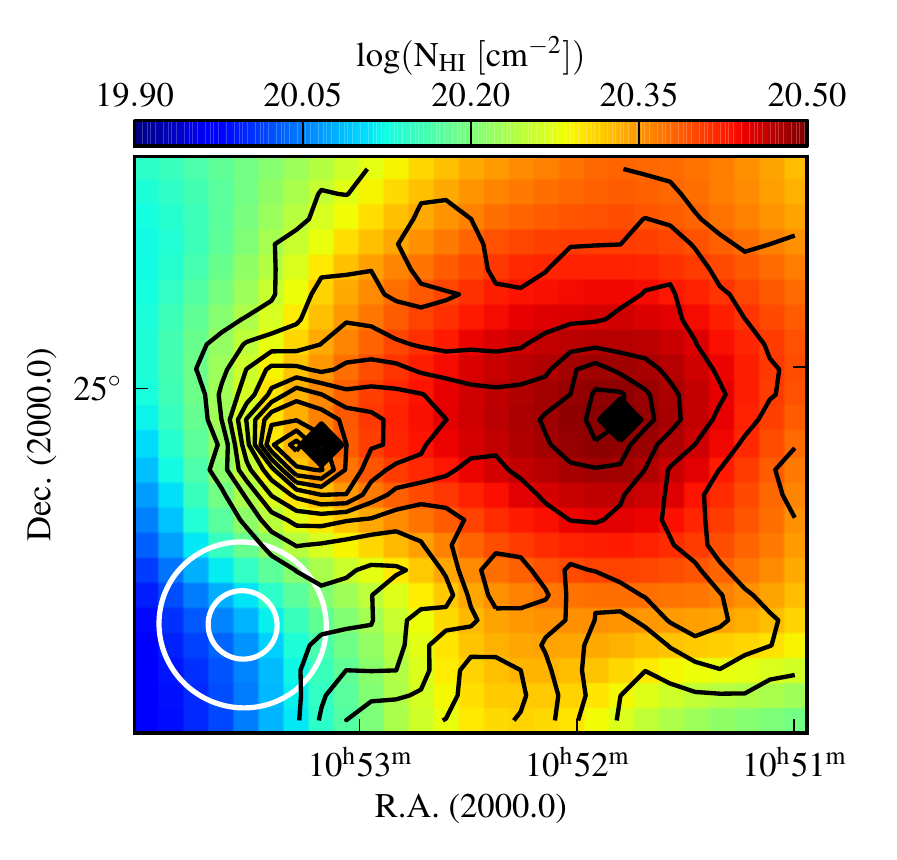}}
  \caption{Column density map of IVC\,2 overlayed with the unsmoothed $857\,\mathrm{GHz}$ data as black contours. A second FIR peak is located east of the $\ion{H}{i}$ core, but it is not evident in $\ion{H}{i}$, implying that this structure is unresolved with our $\ion{H}{i}$ data. The pointing centres of the \citet{Desert1990a} CO observations are given by the two black diamonds. Their corresponding beams are smaller than one pixel. The white circles indicate the angular resolution of EBHIS (larger) and Planck (smaller).}
  \label{fig:ivc2-sub}
\end{figure}

\subsection{Metallicities and distances}
\label{sec:distances-and-metallicities}

In order to derive absolute quantities such as particle densities and cloud masses for the two IVCs, we need a distance estimate. The metallicity is of similar importance for the amount of dust and the formation of molecules.

The cloud IVC\,1 is most likely part of the IV Spur and IVC\,2 of the IV Arch. From absorption spectroscopy, near solar abundances are estimated for the IV Arch and slightly less in the IV Spur \citep{Wakker2001,Richter2001a,Savage1996}. We note that these measurements yield precise values for specific lines of sight only. The two clouds could have different metallicities. However, we expect them to have a comparable metallicity because of their equal FIR brightness at the lowest $N_{\ion{H}{i}}$ (Fig. \ref{fig:ivcs-hi-ir}).

Since there is no accurate distance measurement for either IVC\,1 or IVC\,2, we constrain the distance by several different indicators. Stellar absorption lines restrict the distance to both IV Arch and Spur to the range $0.3 - 2.1\,\mathrm{kpc}$ \citep{Wakker2001}. Similarly, \citet{Puspitarini2012} estimate distances in the range $95-157\,\mathrm{pc}$ to the gas with declinations $\delta \leq 10^{\circ}$ at the bottom of our field. In addition they report on a negative velocity structure between $-40\,\mathrm{km}\,\mathrm{s}^{-1} \leq v_{\mathrm{LSR}} \leq -20 \,\mathrm{km}\,\mathrm{s}^{-1}$ for which they detect no absorption. This IVC gas is at a minimal distance of $200\,\mathrm{pc}$, which leads \citeauthor{Puspitarini2012} to the conclusion that the IVCs are probably not associated with the local $\ion{H}{i}$ shells. A distance estimate for IVC\,2 in the range $12 - 400\,\mathrm{pc}$ is given by \citet{Wesselius1973} based on calcium absorption lines.

To constrain the distance of both IVCs more accurately, we compare the ROSAT soft X-ray shadows \citep{Snowden2000} to the shadow of the molecular IVC\,135+54 for which \citet{Benjamin1996} establish a distance of $355 \pm 95$\,pc by interstellar absorption lines. Because of their high column densities ($N_{\ion{H}{i}} \geq 3 \times 10^{20}\,{\rm cm^{-2}}$), all three IVCs can be considered to be opaque for soft X-rays originating from beyond \citep{Kerp2003}. Hence, the observed soft X-ray count rates towards the IVCs only trace the emission from the galactic foreground plasma.

In the catalogue of soft X-ray shadows compiled by \citet{Snowden2000}, IVC\,1 is listed as cloud 273, IVC\,2 as cloud 241, and our reference cloud IVC\,135+54 as cloud 182. Assuming that for IVC\,135+54 the soft X-ray count rate and its uncertainty fully correspond to the distance estimate of \citet{Benjamin1996}, one can evaluate distances for IVC\,1 and IVC\,2 from their count rates. This yields $D_{\mathrm{IVC\,1}}= 420 \pm 190\,\mathrm{pc}$ and $D_{\mathrm{IVC\,2}}=510 \pm 140\,\mathrm{pc}$. Considering the uncertainties in the ROSAT count rates and the inaccuracies in transforming them into a distance, we adopt the distance estimate $D=0.4\,\mathrm{kpc}$ for both IVC\,1 and IVC\,2. This is compatible with all the other distance indicators we have.

\begin{table*}[!t]
  \caption{Properties of the cores of IVC\,1 and IVC\,2 for a distance of $D=0.4\,\mathrm{kpc}$. The columns give (from left to right) the linear diameter $d$, the upper limit of the gas temperature $T_{\mathrm{kin}}$ for the colder component, the $\ion{H}{i}$ volume density $n_{\ion{H}{i}}$, the maximum pressure $\frac{p_{\mathrm{max}}}{k_{\mathrm{B}}}$, the $\ion{H}{i}$ mass $M_{\ion{H}{i}}$, the dust mass $M_{\mathrm{D}}$, the $\ion{H}{i}$ surface mass density $\Sigma_{\ion{H}{i}}$, the dust surface mass density $\Sigma_{\mathrm{D}}$, the dust optical depth $\tau_{857}^{\mathrm{peak}}$, and dust temperature $T_{\mathrm{D}}^{\mathrm{peak}}$ at the peak $\ion{H}{i}$ column density.}  
  \label{tab:phys-prop-clou}
  \small
  \centering
  \begin{tabular}{ccccccccccc}
    \hline\hline
    IVC & $d$ & $T_{\mathrm{kin}}$ & $n_{\ion{H}{i}}$ & $\frac{p_{\mathrm{max}}}{k_{\mathrm{B}}}$ & $M_{\ion{H}{i}}$ & $M_{\mathrm{D}}$ &  {$\Sigma_{\ion{H}{i}}$} &  {$\Sigma_{\mathrm{D}}$} & $\tau_{857}^{\mathrm{peak}}$ & $T_{\mathrm{D}}^{\mathrm{peak}}$ \\ 
    & [pc] & [K] & [$\mathrm{cm}^{-3}$] & [$\mathrm{K}\,\mathrm{cm}^{-3}$] & [$\mathrm{M}_{\odot}$] & [$\mathrm{M}_{\odot}$] & [$\mathrm{M}_{\odot}\,\mathrm{pc}^{-2}$] & [$\mathrm{M}_{\odot}\,\mathrm{pc}^{-2}$] & & [$\mathrm{K}$] \\
    \hline
    1 & $19$ & $441$ & $7$ & $2880$ & $620$ & $2.1$ & $2.1$ & $0.0073$ & ($8.3 \pm 0.4) \times 10^{-6}$ & $21.2 \pm 0.2$ \\
    2 & $10$ & $567$ & $10$ & $5890$ & $140$ & $1.2$ & $1.9$ & $0.0163$ & ($3.2 \pm 0.1) \times 10^{-5}$ & $19.1 \pm 0.2$ \\
    \hline
  \end{tabular}
\end{table*}

\begin{figure*}[!t]
  \centering
  \subfloat{\includegraphics[width=0.5\textwidth]{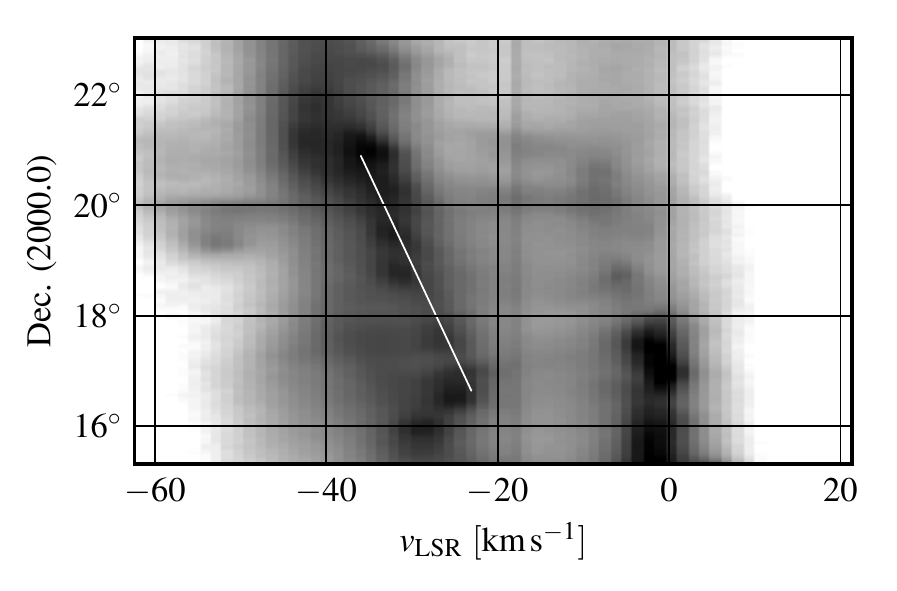}}
  \subfloat{\includegraphics[width=0.5\textwidth]{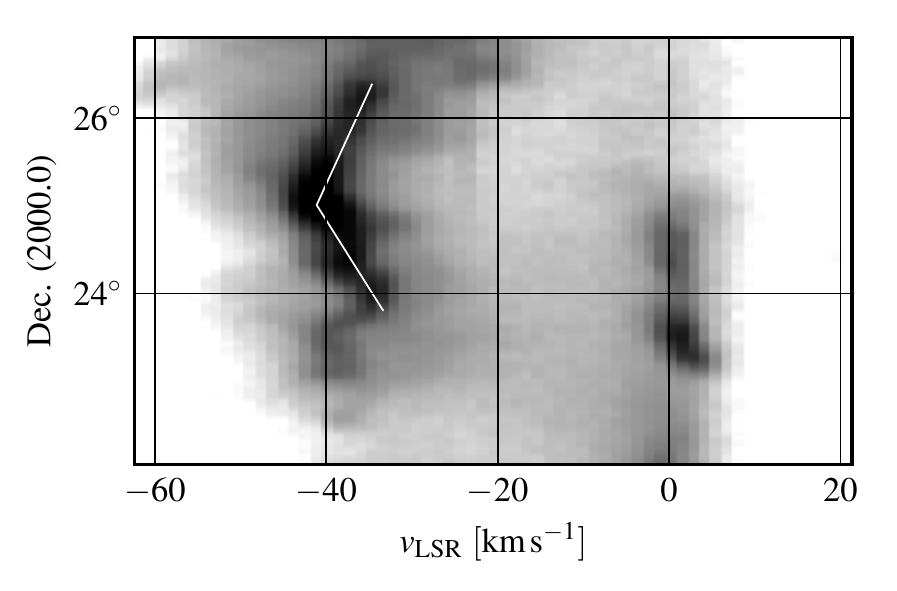}}\\
  \caption{Position-velocity (PV) diagrams of IVC\,1 (\textit{left}) integrated between $11^{\mathrm{h}}57^{\mathrm{m}} \geq \mathrm{R.A.} \geq 11^{\mathrm{h}}45^{\mathrm{m}}$ and IVC\,2 (\textit{right}) integrated between $10^{\mathrm{h}}54^{\mathrm{m}} \geq \mathrm{R.A.} \geq 10^{\mathrm{h}}48^{\mathrm{m}}$. The white lines indicate the orientation of the IVCs in PV space showing a velocity gradient which is interpreted as evidence for interactions of the clouds with the surrounding medium.}
  \label{fig:ivcs-pv}
\end{figure*}

\subsection{Estimation of cloud parameters}
\label{sec:estimation-of-cloud-parameters}

To characterise the clouds more completely, we estimate the following parameters:
\begin{itemize}
\item  An upper limit on the kinetic gas temperature $T_{\mathrm{kin}}$ is obtained from the $\ion{H}{i}$ line width $\Delta v$ \citep[their Eq. 4]{Kalberla2009}. In addition to Doppler broadening, turbulence and substructure also contribute to the spectral line.

\item The physical size $d$ is estimated from the angular size and the distance $D$, assuming a circular cloud.
\item The volume density $n_{\ion{H}{i}}$ is considered to be constant inside a spherically symmetric cloud of physical size $d$.
\item The $\ion{H}{i}$ mass $M_{\ion{H}{i}}$ of a cloud is estimated by spatially integrating over $N_{\ion{H}{i}}$ given an estimate for the distance $D$.

\item According to \citet{Hildebrand1983}, the dust mass $M_{\mathrm{D}}$ of a cloud is calculated from the observed FIR brightness by 
\begin{equation}
\label{eq:3}
 M_{\mathrm{D}} = \frac{I_{\nu} \Omega D^{2}}{B_{\nu}(T_{\mathrm{D}})} \times \frac{\frac{4}{3}a\rho_{\mathrm{D}}}{Q_{\nu}} \simeq \frac{I_{857} \Omega D^{2}}{B_{857}(T_{\mathrm{D}})} \times 0.21\,\mathrm{g}\,\mathrm{cm}^{-2}
\end{equation}
with the FIR intensity $I_{\nu}$, the solid angle $\Omega$ of the cloud, the distance $D$, the grain radius $a$, the grain emissivity $Q_{\nu}$, and the grain density $\rho_{\mathrm{D}}$. We estimate the dust mass for $857\,\mathrm{GHz}$ with the values for $a$, $Q_{\nu}$, and $\rho_{\mathrm{D}}$ given by \citeauthor{Hildebrand1983}.
\end{itemize}

Many cloud parameters depend on the distance, for which we use $D = 0.4\,\mathrm{kpc}$ (Sect. \ref{sec:distances-and-metallicities}). We calculate the physical quantities for the cores which are marked by the white contour in Figs. \ref{fig:ivc1-int} and \ref{fig:ivc2-int}. We use a watershed algorithm \citep{Beucher1979} to determine the extent of these cores.

The results for the parameters are compiled in Table \ref{tab:phys-prop-clou}. We note that for IVC\,2 no molecular hydrogen is taken into account; H$_{2}$ adds to the total particle density and the total gas mass.

For the $\ion{H}{i}$ surface mass density, we obtain for the IVC cores $\Sigma_{\ion{H}{i}}^{\mathrm{IVC\,1}}\simeq 2.1 \,\mathrm{M}_{\odot}\,\mathrm{pc}^{-2}$ and $\Sigma_{\ion{H}{i}}^{\mathrm{IVC\,2}}\simeq 1.9 \,\mathrm{M}_{\odot}\,\mathrm{pc}^{-2}$. We calculate a corresponding dust surface mass density which results in $\Sigma_{\mathrm{D}}^{\mathrm{IVC\,1}}\simeq 0.0073 \,\mathrm{M}_{\odot}\,\mathrm{pc}^{-2}$ and $\Sigma_{\mathrm{D}}^{\mathrm{IVC\,2}}\simeq 0.0163 \,\mathrm{M}_{\odot}\,\mathrm{pc}^{-2}$. The $\ion{H}{i}$ surface density is comparable, but in IVC\,2 the dust surface density is more than twice as large. 

Both IVC\,1 and IVC\,2 are not self-gravitating. Their Jeans masses, derived from the $\ion{H}{i}$ data, are larger than their gas masses by two orders of magnitude. This is still true when H$_{2}$ is considered. However, the gas temperature estimated from the $\ion{H}{i}$ serves only as an upper limit. Locally, the gas is certainly colder when we consider the CO within IVC\,2, possibly even below $30\,\mathrm{K}$ \citep[see e.g.][]{Glover2012}. Nevertheless, we do not expect IVC\,2 to form stars.

\subsection{Interactions of IVCs with the ambient medium}
\label{sec:vel-gra}

Indications for dynamical interactions of IVC\,1 and IVC\,2 with their environment are inferred from position-velocity (PV) diagrams (Fig. \ref{fig:ivcs-pv}) that show the velocity changes over the IVC filaments. The PV diagrams are integrated between $11^{\mathrm{h}}57^{\mathrm{m}} \geq \mathrm{R.A.} \geq 11^{\mathrm{h}}45^{\mathrm{m}}$ for IVC\,1 and $10^{\mathrm{h}}54^{\mathrm{m}} \geq \mathrm{R.A.} \geq 10^{\mathrm{h}}48^{\mathrm{m}}$ for IVC\,2. The white diagonal lines in the diagrams mark the location and orientation of the two IVCs in PV space, revealing a change of $v_{\mathrm{LSR}}$ over the IVCs and the filaments to which they are connected. These gradients are probably due to interactions of the clouds with their surroundings. The PV-diagrams also reveal the clumpy structure along the IVC filaments.

The isothermal speed of sound in an ideal gas with the kinetic temperature of IVC\,1 or IVC\,2 is about $2.0\,\mathrm{km}\,\mathrm{\mathrm{s}}^{-1}$ for the colder components. Thus the observed radial LSR velocity gradient within the individual IVCs significantly exceeds the speed of sound in the cold neutral medium. If this gradient is physical and not the result of projection effects, it implies that the IVC cores are punching through the galactic halo medium while material is being stripped off and decelerated. A supersonic deceleration would build up shocks that compress and fragment the clouds \citep[e.g.][]{McKee1980}.

\section{Discussion: a dynamical $\ion{H}{i}$-H$_{2}$ transition}
\label{sec:discussion}

We observe two IVCs in close proximity to each other that show respectively a deficiency (IVC\,1) and an excess (IVC\,2) in FIR emission, despite their similarities in $\ion{H}{i}$ and in their environmental conditions:
\begin{itemize}
\item Both IVCs consist of distinct cold clumps marked by narrow spectral $\ion{H}{i}$ lines with $\mathrm{FWHM}\simeq5\,\mathrm{km}\,\mathrm{\mathrm{s}}^{-1}$. Their $\ion{H}{i}$ profiles can be characterised by this cold component, plus a warmer component with $\mathrm{FWHM}\simeq11\,\mathrm{km}\,\mathrm{\mathrm{s}}^{-1}$. About $50\%$ of the $\ion{H}{i}$ emission of IVC\,1 originates from the cold gas, whereas for IVC\,2 $70\%$ is in the cold phase. It is remarkable that for IVC\,1 we estimate $N_{\ion{H}{i}} \simeq 3.9 \times 10^{20}\,\mathrm{cm}^{-2}$, which is significantly larger than for IVC\,2 with $N_{\ion{H}{i}} \simeq 3.1 \times 10^{20}\,\mathrm{cm}^{-2}$ (Sects. \ref{sec:ivc1}, \ref{sec:ivc2}).
\item To constrain the distances to both IVCs, we compare the ROSAT soft X-ray shadows of the IVCs to the molecular IVC\,135+54 for which there is a firm distance bracket. We conclude that IVC\,1 and IVC\,2 have a similar distance similar to IVC\,135+54 of about $D=0.4\,\mathrm{kpc}$ (Sect. \ref{sec:distances-and-metallicities}).
\item Both IVCs show an equal $\ion{H}{i}$ surface mass density. However, the dust surface mass density of IVC\,2 is more than twice as large as that of IVC\,1 (Sect. \ref{sec:estimation-of-cloud-parameters}).
\item Both IVCs have radial velocities between $-35\,\mathrm{km}\,\mathrm{s}^{-1}$ and $-40\,\mathrm{km}\,\mathrm{s}^{-1}$. They exhibit a velocity gradient of $10\,\mathrm{km}\,\mathrm{s}^{-1}$ (Sect. \ref{sec:vel-gra}).
\item From the FIR excess emission we estimate molecular column densities for IVC\,2 of $N_{\mathrm{H}_{2}}=1-2\times10^{20}\,\mathrm{cm}^{-2}$ which gives a total molecular fraction of $f^{\mathrm{IVC\,2}}_{\mathrm{H}_{2}}\simeq0.4$ (Sect. \ref{sec:map-of-mol-hyd-col-den}).
\end{itemize}
In order to reconcile the similarities in $\ion{H}{i}$ properties with the low FIR emission of IVC\,1 and the excess emission of IVC\,2 due to H$_{2}$, we propose that IVC\,1 and IVC\,2 represent different states in a phase transition from atomic to molecular clouds at the disk-halo interface. The descent of IVCs onto the galactic disk is thought to compress the gas which increases the pressure locally, triggering the fast formation of H$_{2}$ \citep{Odenwald1987,Desert1990a,Weiss1999,Gillmon2006a,Guillard2009}.

\subsection{The scenario of interacting IVCs}
\label{sec:the-scenario-of-impacting-ivcs}

According to the galactic fountain model, metal enriched disk material is ejected via supernovae into the disk--halo interface \citep{Bregman2004}. This rising gas is warm and ionised as a result of the large energies involved in the expulsion.

After culmination, the ejected matter falls back onto the disk. The descent from the culmination point is accompanied by an increase in gas pressure due to ram pressure. Electrons and protons recombine and form the warm neutral medium (WNM). When the ram pressure becomes larger than the thermal pressure within the cloud, shocks are induced that propagate through the IVC. These shocks enhance the pressure locally by which the formation time of H$_{2}$ is decreased \citep{Guillard2009}. The shocks trigger cooling of the WNM into condensations of the cold neutral medium (CNM) where H$_{2}$ forms.

The ram pressure not only compresses the cloud, but also causes a deceleration. This has been modelled by \citet{Heitsch2009}. Eventually, the IVC may reach local velocities which make it indistinguishable from local clouds. Since there are only a few molecular IVCs (IVMCs) known \citep{Magnani2010}, a certain fine tuning of the parameters seems necessary in order to create an IVMC that we can detect.

In the following we try to evaluate the scenario of ram pressure induced H$_{2}$ formation by looking at the pressures and the timescales involved in the cases of IVC\,1 and IVC\,2.

\subsection{H$_{2}$ formation in compressed gas}
\label{sec:h2-formation}

\citet{Bergin2004} write the time evolution of the H$_{2}$ number density $n_{\rm{H}_{2}}$ as
\begin{equation}
 \label{eq:nh2}
 \frac{\mathrm{d}n_{\mathrm{H}_{2}}}{\mathrm{d}t}=R_{\mathrm{gr}}(T)n_{\mathrm{H}_{2}}n_{\mathrm{H}} - [\zeta_{\mathrm{cr}}+\zeta_{\mathrm{diss}}(N_{\mathrm{H}_{2}},A_{\mathrm{V}})]n_{\mathrm{H}_{2}}
\end{equation}
with the temperature dependent grain formation rate $R_{\mathrm{gr}}(T)$, the destruction rate by cosmic rays $\zeta_{\mathrm{cr}}$, and the photo-dissociation rate $\zeta_{\mathrm{diss}}(N_{\mathrm{H}_{2}},A_{\mathrm{V}})$. The dissociation rate $\zeta_{\mathrm{diss}}$ depends on the self-shielding of H$_{2}$ \citep{Draine1996}.

Equation \eqref{eq:nh2} relates the formation time of H$_{2}$ to the particle density and hence the pressure. We refer to the late stages of the infall during which the IVC has accumulated particle densities $n_{\ion{H}{i}}>1\,\mathrm{cm}^{-3}$. Compressed gas is able to cool quickly (Sect. \ref{sec:time-scales}) which results in large particle densities.

\begin{figure}[!t]
 \centering
  \resizebox{\hsize}{!}{\includegraphics[width=1.\textwidth]{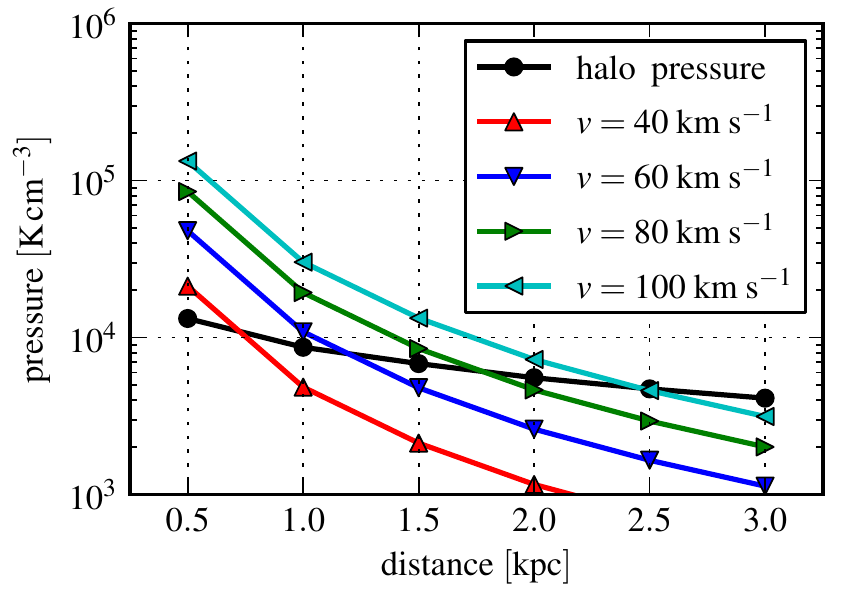}}
  \caption{Comparison of modelled halo pressure from \citet{Kalberla2008} to ram pressure at the given distance in the line of sight towards IVC\,2. The ram pressure is calculated from Eq. \eqref{eq:ram-p} using the modelled halo density and the chosen cloud velocities $40$, $60$, $80$, and $100\,\mathrm{km}\,\mathrm{s}^{-1}$.}
  \label{fig:halo-ram-press}
\end{figure}

In our picture ram pressure acts on the cloud because of its movement through the ISM. The ram pressure
\begin{equation}
\label{eq:ram-p}
p_{\mathrm{ram}} = \rho \times v^{2}
\end{equation}
depends on the mass density $\rho$ of the surrounding medium and the relative velocity $v$ between cloud and surroundings. The cloud moves through the galactic halo, for which we take the Milky Way model of \citet{Kalberla2008} to get estimates of the halo densities and pressures. We calculate the associated ram pressure for several velocities at the distances $0.5\,\mathrm{kpc} \leq D \leq 3\,\mathrm{kpc}$ in the line of sight towards IVC\,2. In Fig. \ref{fig:halo-ram-press} we compare the ram pressure with the halo pressure. We note that we can observe only the radial velocity component. Any additional tangential velocity component is unknown.

At one point the ram pressure exceeds the halo pressure for all considered cloud velocities. The distance at which the ram pressure and the halo pressure are equal increases with cloud velocity. Thus, ram pressure interactions are expected to be important for all IVCs. For IVC\,1 and IVC\,2, which are at a distance of $D\simeq0.4\,\mathrm{kpc}$ (Sect. \ref{sec:distances-and-metallicities}), the ram pressure is at least twice as large as the halo pressure. Hence, perturbations driven by ram pressure are not only possible, but likely. Indications for this are for example the morphology of the clouds (Sects. \ref{sec:ivc1} and \ref{sec:ivc2}) and their velocity gradients (Sect. \ref{sec:vel-gra}).

We emphasise the importance of magnetic fields for cloud compression and condensation. In the galactic halo there are magnetic fields of a few $\mu \mathrm{G}$ which can be important for gas dynamics \citep{Putman2012}. \citet{Hartmann2001} show that clouds can form preferentially at kinks or bends in the magnetic field.

\subsection{Timescales}
\label{sec:time-scales}

From isobaric cooling models \citet{Guillard2009} find an inverse scaling between the H$_{2}$ formation timescale and the pressure which can be approximated by
\begin{equation}
\label{eq:th2}
t_{\mathrm{H}_{2}}[\mathrm{yr}]\simeq 7 \times 10^{5} f_{\mathrm{dust}} \left( \frac{2\times 10^{5}\,[\mathrm{K}\,\mathrm{cm}^{-3}]}{p_{\mathrm{th}}} \right) ^{0.95}.
\end{equation}
We note that the H$_{2}$ formation timescale in Eq. \eqref{eq:th2} is computed neglecting H$_{2}$ destruction in Eq. \eqref{eq:nh2}. \citet{Bergin2004} give full solutions to Eq. \eqref{eq:nh2} with longer timescales by up to a factor of two. In our case $f_{\mathrm{dust}}=1$ since the emissivity per H atom from IVC\,2 is comparable to the mean value in the low velocity gas. 

For a distance $D\simeq0.4\,\mathrm{kpc}$ the ram pressure in both IVC\,1 and IVC\,2 is $p_{\mathrm{ram}}\geq2\times10^{4} \, \mathrm{K}\,\mathrm{cm}^{-3}$, resulting in a formation timescale of $t_{\mathrm{H}_{2}}\lesssim 6\,\mathrm{Myr}$. In this time the clouds travel $\lesssim 0.25\,\mathrm{kpc}$. \citet{Bergin2004} state that the actual formation timescale of H$_{2}$ is not the limiting factor in molecular cloud formation; it is merely a requirement of shielding of H$_{2}$ and CO which is governed by the accumulation of gas and dust and the formation of dense cores. Hence, $t_{\mathrm{H}_{2}}$ could be even smaller.

For the CNM, \citet{Kalberla2009} estimate a cooling time of $t_{\mathrm{cool}}(\mathrm{CNM})\simeq0.1\,\mathrm{Myr}$. This is significantly less than $t_{\mathrm{H}_{2}}$, explaining why ISM clouds contain CNM without H$_{2}$. Above a sufficient particle density, compressed gas can cool faster and become denser.

To estimate the time an IVC has to form H$_{2}$, we need to compare the formation time with the free-fall time in the galactic gravitational potential calculated by applying a simple linear, unaccelerated motion. Assuming an initial velocity of $v=80\,\mathrm{km}\,\mathrm{s}^{-1}$, the free-fall time from an altitude of $0.5\,\mathrm{kpc}$ in the halo is about $6\,\mathrm{Myr}$. Thus, the H$_{2}$ formation timescale is comparable or shorter than the descending time; H$_{2}$ may form before the IVC merge with gas in the galactic plane. 

It is not surprising that IVC\,1 and IVC\,2 are not different with respect to their estimated H$_{2}$ formation timescale, since $t_{\mathrm{H}_{2}}$ is inferred from the EBHIS data only, in which both clouds are very similar (Sect. \ref{sec:individual-clouds}). Hence, the observed differences in the FIR are a result of processes on angular scales which are unresolved with EBHIS.

\subsection{Molecular gas in the field}
\label{sec:nolecular-gas-in-the-field}

For the molecular IVC we sketch a scenario of H$_{2}$ formation by the compression due to ram pressure reducing the H$_{2}$ formation timescale. However, two major questions remain: will IVC\,1 become molecular in future? and how is the H$_{2}$ at $v_{\mathrm{LSR}}=0\,\mathrm{km}\,\mathrm{s}^{-1}$ formed?

At present we consider the atomic IVC\,1 to be in an intermediate state of molecular formation. Either it is already forming H$_{2}$ efficiently in small condensations within the cloud, or it will in the future.

In the position-velocity (PV) diagrams of IVC\,1 and IVC\,2 (Fig. \ref{fig:ivcs-pv}) both clouds are clearly connected to lower velocity gas. Furthermore, the observed velocity gradients along the IVC filaments may indicate a dynamical connection of IVC to lower velocity gas. It is tempting to speculate that some H$_{2}$ in the local gas represents the remnants of a past interaction between IVC gas and the galactic disk. As IVC gas slows down, the ram pressure and the H$_{2}$ formation rate decrease. Once a sufficiently high amount of H$_{2}$ has been formed, self-shielding becomes important and maintains the large H$_{2}$ content.

The outlined dynamically--triggered formation of H$_{2}$ is not necessarily traceable by CO emission and might be considered as CO-dark gas \citep{Wolfire2010,Planckcollaboration2011XIX}. In particular, a sufficiently high dust column density has to be built up in order to allow the formation of CO. Related to this is the X$_{\mathrm{CO}}$ factor which is known to be influenced by local conditions especially by metallicity \citep{Feldmann2012}. Furthermore, CO has to be collisionally excited to become observable in emission. \citet{Bergin2004} point out that there could be a large reservoir of H$_{2}$ in the diffuse ISM which is not able to form CO detectable in emission.

\section{Conclusion}
\label{sec:conclusion}

We correlate $\ion{H}{i}$ emission to the brightness at various wavelengths in the FIR dust continuum using new data from the  Effelsberg-Bonn $\ion{H}{i}$ Survey (EBHIS) and the Planck satellite complemented by IRIS data. We study in detail two IVCs that show many similarities in their $\ion{H}{i}$ properties, such as narrow spectral lines with $\mathrm{FWHM}\simeq5\,\mathrm{km}\,\mathrm{\mathrm{s}}^{-1}$ and $\ion{H}{i}$ column densities of $N_{\ion{H}{i}} = 3 - 4 \times 10^{20}\,\mathrm{cm}^{-2}$. Despite their similarity in $\ion{H}{i}$, their FIR emission exhibits large differences: one cloud is FIR bright while the other IVC is FIR faint.

From the quantitative correlation of $\ion{H}{i}$ and FIR emission, we calculate maps of molecular hydrogen column density revealing large amounts of H$_{2}$ in the field of interest for which no existing CO surveys of the region has detected a CO counterpart. How much of this H$_{2}$ is actually CO-dark gas we cannot tell since the CO survey data that is available today is not sensitive enough. We do, however, know that the molecular IVC contains CO. The $\ion{H}{i}$ emission traces only a part of the total gas distribution. Together with the inferred H$_{2}$ column densities, the relation between gas and dust is consistent.

Based on our findings we describe a scenario of a dynamical transition from atomic to molecular IVCs in the lower galactic halo. During the descent of the IVCs through the galactic halo, they are compressed as a result of external pressure and ram pressure. Once the ram pressure exceeds the thermal pressure, shocks are created which enhance the pressure locally and accumulate gas and dust. An increased pressure reduces the formation timescale of H$_{2}$ in condensations of the cold atomic medium. 

The only physical distinctions between the two IVCs are the amount of dust within the clouds, measured by the dust surface mass density, and the amount of H$_{2}$. The molecular IVC has a factor of $2-3$ more dust within its central region than the atomic IVC, which is accompanied by a $50\%$ higher total hydrogen column density. On the other hand, IVC\,1 has more $\ion{H}{i}$ mass in total. Apparently, this mass is not distributed so as to allow efficient H$_{2}$ formation. According to the data presented in this paper, we expect that the atomic IVC will also turn molecular in a few Myrs.

Processes on spatial scales that are not resolved by our data govern the evolution of an IVC from an atomic to a molecular cloud. The $10.8\,\arcmin$ resolution of EBHIS corresponds to a spatial resolution of $1\,\mathrm{pc}$ at a distance of $0.4\,\mathrm{kpc}$. However, the accumulation and condensation of smaller and denser clumps regulate the H$_{2}$ formation on sub-parsec scales. Radio-interferometric observations should reveal a different distribution of $\ion{H}{i}$ in the two IVCs on sub-parsec scales, for example compact cores in the molecular IVC.

Our approach appears to open a way to search for dark H$_{2}$ gas across the entire sky. Globally, this new search may reveal other clouds in transition from the atomic to the molecular gas phase. Compared to the low angular resolution of former large-scale $\ion{H}{i}$ single-dish surveys, the few detections of molecular IVCs so far may be due to the small-angular extent of the molecular cores.

\begin{acknowledgements}
We thank the anonymous referee for his useful comments and suggestions which helped to improve the manuscript considerably. The authors thank the Deutsche Forschungsgemeinschaft (DFG) for financial support under the research grant KE757/11-1. F. B. acknowledges support from the MISTIC ERC grant no. 267934. The work is based on observations with the 100$\,$m telescope of the MPIfR (Max-Planck-Institut f{\"u}r Radioastronomie) at Effelsberg and the Planck satellite operated by the European Space Agency. The development of Planck has been supported by: ESA; CNES and CNRS/INSU-IN2P3-INP (France); ASI, CNR, and INAF (Italy); NASA and DoE (USA); STFC and UKSA (UK); CSIC, MICINN and JA (Spain); Tekes, AoF and CSC (Finland); DLR and MPG (Germany); CSA (Canada); DTU Space (Denmark); SER/SSO (Switzerland); RCN (Norway); SFI (Ireland); FCT/MCTES (Portugal); and PRACE (EU). T. R. is a member of the International Max Planck Research School (IMPRS) for Astronomy and Astrophysics at the Universities of Bonn and Cologne 
as well as of the Bonn-Cologne Graduate School of Physics and Astronomy (BCGS). 
\end{acknowledgements}

%
%

\bibliographystyle{aa} 
\bibliography{Literatur} 

\end{document}